\documentclass[fleqn,usenatbib]{mnras}

\usepackage{newtxtext,newtxmath}

\usepackage[T1]{fontenc}

\DeclareRobustCommand{\VAN}[3]{#2}
\let\VANthebibliography\thebibliography
\def\thebibliography{\DeclareRobustCommand{\VAN}[3]{##3}\VANthebibliography}

\usepackage{graphicx}	
\usepackage{amsmath}	

\usepackage{amssymb}	

\usepackage{threeparttable}

\title[ELM WD Binary]{Orbital parameters for an ELM white dwarf with a white dwarf companion: LAMOST J033847.06+413424.2}

\author[Yuan et al.]{
Hailong Yuan,$^{1}$\thanks{E-mail: yuanhl@bao.ac.cn}
Zhenwei Li,$^{2}$
Zhongrui Bai,$^{1}$
Yiqiao Dong,$^{1}$
Yao Cheng,$^{3}$
Xuefei Chen,$^{2}$
\newauthor
Zhixiang, Zhang, $^{2}$
Mengxin Wang,$^{1}$
Mingkuan Yang,$^{1}$
Xin Huang,$^{1}$
Yuji He,$^{1}$
Liyun Zhang,$^{4}$
\newauthor
Junfeng Wang,$^{3}$
Yongheng Zhao,$^{1}$
Yaoquan Chu $^{5}$
and Haotong Zhang,$^{1}$\thanks{E-mail: htzhang@bao.ac.cn}
\\
$^{1}$Key Laboratory of Optical Astronomy, National Astronomical Observatories, Chinese Academy of Sciences, Beijing 100101, China\\
$^{2}$Yunnan Observatories, Chinese Academy of Sciences, 650011, China\\
$^{3}$Department of Astronomy, Xiamen University, Xiamen, Fujian 361005, China\\
$^{4}$Department of Physics and Astronomy, Guizhou University, Guiyang, Guizhou 550025, China\\
$^{5}$University of Science and Technology of China, Hefei 230026, China
}

\date{Accepted XXX. Received YYY; in original form ZZZ}

\pubyear{2023}

\begin{document}
\label{firstpage}
\pagerange{\pageref{firstpage}--\pageref{lastpage}}
\maketitle

\begin{abstract}
Double white dwarf systems are of great astrophysical importance in the field of gravitational wave and Type Ia supernova.
While the binary fraction of CO core white dwarf is about a few percents, 
the extremely low mass white dwarfs are all thought to be within binary systems.
In this work, we report the orbital solution of a double degenerate system: J033847.06+413424.24,
an extremely low mass He core white dwarf orbiting a CO core white dwarf.
With LAMOST and P200, time domain spectroscopic observations have been made
and spectral atmosphere parameters are estimated to be $T_{\rm eff}\sim22500$ K and log $g\sim5.6$ dex.
Combing Gaia parallax, 3D extinction, and evolution tracks, 
we estimate a radius of $\sim0.12$ $R_{\odot}$ and a mass of $\sim0.22$ $M_{\odot}$.
With the 37 single exposure spectra, the radial velocities are measured 
and the orbital parameters are estimated to be $P=0.1253132(1)$ days, $K1=289\pm4$ km/s and $V_{sys}=-41\pm3$ km/s.
The radial velocity based system ephemeris is also provided.
The light curves from several photometric surveys show no orbital modulation.
The orbital solution suggests that 
the invisible companion has a minimum mass of about 0.60 $M_{\odot}$ 
and is $\sim0.79$ $M_{\odot}$ for an inclination of $60.0^{\circ}$,
indicating most probably a CO core white dwarf.
The system is expected to merge in about 1 Gyr.
With present period and distance ($\sim596$ pc) it can not irradiate 
strong enough gravitational wave for LISA.
More double degenerate systems are expected to be discovered and parameterized as the LAMOST survey goes on.
\end{abstract}




\section{INTRODUCTION}
\label{intro.sec}


Double white dwarf (DWD) systems consist of two white dwarfs (WDs).
While single WDs will cool down quietly following the WD cooling sequence,
DWD systems are related to many interesting phenomena and can provide an excellent laboratory for studying stellar physics.
As the two components of DWDs spiral towards each other, 
gravitational wave (GW) is emitted and can be detected by sensitive gravitational wave detectors such as 
LISA \citep[][]{Amaro2023LISA}, 
Virgo \citep[][]{Acernese2015VIRGO},
GEO 600 \citep[][]{Grote2010},
KAGRA \citep[][]{Kagra2019},
TAIJI \citep[][]{Zhong2023Taiji}
and TianQin \citep[][]{Huang2020TianQin}.
DWDs are important as they have the largest number in the mHz frequency range
\citep[][]{Korol2022DD,lizw2023}.
DWDs also represent one of the most important formation pathways of Type Ia Super Nova (Ia SN),
which is important cosmological distance indicator.
Understanding and optimizing the Ia SN forming theory is critical to better constrain their luminosities and distances \citep[][]{Wang2012IaSN,Liuz2023}.
The DWD channel is one of the three AM CVn formation pathways \citep[][]{Solheim2010AmCVn},
although \citet[][]{Brown2016} argued that DWDs are progenitors of all AM CVn binaries
based on the ELM space density analysis.
Low mass DWDs will eventually merge into single massive white dwarfs \citep[][]{Brown2016}.

The DWD population is expected to be rich,
but the total number,
especially for systems with orbital solutions,
is still small, 
due to the fact that time domain spectroscopic observation is fairly time consuming.
The ESO/SPY project, 
which is aimed to search for DWDs \citep[][]{Nelemans2005},
has lead to a white dwarf binary fraction of about 5.7\% (39 out of 679 systems) \citep[][]{Koester2009, Napiwotzki2020}.
Based on the Palomar-Green survey, 
\citet[][]{Brown2011} made a sample of 30 low mass White dwarfs ($\leq0.45$ $M_{\odot}$),
and suggested a binary fraction of at least 70\%, 
although the nature of their companions are unambiguously constrained.
With a limit distance of 25 pc, 
\citet[][]{Holberg2016} has presented a sample of 232 white dwarfs,
out of which there are 24 ($\approx10\%$) WD$+$WD systems.
The Sloan Digital Sky Survey (SDSS) has lead to 
the discoveries of a few DWDs 
\citep[][]{Badenes2009SDSS1257, Kulkarni2010SDSS1257, Mullally2009DD, Kilic2010, Chandra2021SDSS1337}.
As summarized by \citet[][]{Korol2022DD} that there are less than 150 DWD systems with known orbital solution.
Recently the Extremely Low Mass (ELM) white dwarf survey has already discovered nearly 150 unique ELM binaries \citep[][]{Brown2020ELM, Brown2022ELM, Kosakowski2020ELM, Kosakowski2023ELM},
suggesting a very high double degenerate fraction (almost 100\%).

As a hot subdwarf $+$ WD system will evolve into a double degenerate system within 100 million years \citep[][]{Heber2016subdwarf},
it can be counted as a DWD progenitor.
After the PG survey about 7 subdwarf+WD binaries were found \citep[][]{Saffer1998}.
The following ESO/SPY project suggested a high subdwarf binary fraction of about $\sim50\%$ \citep[][]{Morales2003subdwarf, Napiwotzki2004}.
\citet[][]{Kupfer2015sdB} summarised a sample of 142 hot subdwarf binaries from MUCHFUSS project,
and about a half have white dwarf companions.

Technically, time domain spectroscopic observation upon color selected sample 
is still the most efficient way to find and study DWDs.
Recently, thanks to the Gaia survey \citep[][]{2018gaia},
an unprecedented catalogue of about 500 thousand white dwarfs has been reported.
\citet[][]{Tremblay2020GaiaWD40pc1} has started a project to extend a spectroscopy survey for a 40 pc white dwarfs sample,
which has a size above 1000, 
and will be the benchmark sample for many years to come \citep[][]{McCleery2020GaiaWD40pc2, OBrien2023GaiaWD40pc3}.
A 100pc sample, with a sample size of about 10 thousand, has also been studied using Gaia data \citep[][]{Torres2019GaiaWD100pc1, Torres2022GaiaWD100pc2, Jim2023GaiaWD100pc3}.
Therefore,
although the white dwarf binary fraction is low,
there are still many unknown DWDs to be discovered by time domain spectral observations.




%

In this work, we present the orbital solution for LAMOST J033847.06+413424.24 (J0338 for short).
J0338 was reported as an ELM candidate 
by both Gaia color magnitude selection method \citep[][]{Pelisoli2019GaiaELM}
and LAMOST spectra \citep[][]{Bai2017, Bai2021, Wang2022ELM}.
Nevertheless, several works also classified J0338 as a hot subdwarf (candidate) \citep[][]{Lei2018sdB, Geier2019subdwarf, Luo2021sdB, Culpan2022subdwarf}.
Both types have experienced significant mass loss by their close companions,
making them close relatives in evolutionary path.
However, a hot subdwarf is under core helium burning state,
while an ELM isn't and instead has a burning hydrogen envelope.
Observationally a hot subdwarf is hotter (separated by $\sim25000$ K) 
and more massive than an ELM (hot subdwarfs are considered to have mass above $0.3$ $M_{\odot}$).

It is significant to correctly distinguish between these two types,
as they have different evolutionary channels \citep[][]{Heber2016subdwarf, Istrate2016ELMs, Li2019ELM} 
and can be used to constrain many unclear physical processes within close binaries, 
like mass transfer, angular momentum lose, common envelope efficiency and et al.
\citep[][]{Nelemans2010, Ostensen2011, Brown2016, Vos2019RL, Ge2022CE, lizw2023}.
Since most of these sources are confirmed spectroscopically,
using spectra from large sky low resolution survey \citep[][]{Brown2022ELM, Geier2022},
it will be of great importance to test if it is possible to make a reliable classification,
and if there are any special data reduction steps need to be treated carefully.
Thus the case study of J0338 here will be instructive to maximize the use of spectral data in other similar works.
The lifetimes of hot subdwarfs are about 100 million years \citep[][]{Heber2016subdwarf},
while ELMs can maintain the hydrogen shell burning stage for 0.5 to 2.0 Gyrs,
depending on their masses \citep[][]{istrate2014}.
Thus it is possible that more ELMs are to be discovered in the future.


We will describe the spectroscopic and photometric observations in Section \ref{sect:obs},
and present the data analysis in Section \ref{sect:result}.
Results and discussions are in Section \ref{sect:summary}.
For clarity, 
we name the visible star which show the orbital radial velocity variation as the primary, or star 1,
and the invisible compact star as the secondary, or star 2.
Furthermore, the orbital phase $\phi = 0$ is defined as the time 
when the primary star is at superior conjunction point.

\section{Observations}
\label{sect:obs}

\subsection{LAMOST Observation}
\label{sect:lamost}

With a 4 meter effective aperture and 4000 fiber, 
Large Sky Area Multi-Object Fiber Spectroscopic Telescope (LAMOST) is a powerful large sky spectroscopic survey telescope 
\citep{1996ApOpt..35.5155W, 2004ChJAA...4....1S}.
By 2022, over 20 million spectra has been released, mainly for celestial targets with magnitudes between 14 and 17, covering 20 thousand square degrees above Declination -10$^{\circ}$.
Target J0338 was visited by LAMOST low resolution survey for three nights in 2015 and 2016, 
corresponding to 9 single exposures, with details shown in Table \ref{tab.sp_obs}.
All these spectra cover wavelength range of 3700 to 9100 {\AA}, 
with an average resolution of $\approx1800$ and signal to noise ratio (S/N) of about 40.
The observed data is processed by LAMOST pipeline \citep{Bai2017, Bai2021}.

\subsection{P200 Observation}
\label{sect:p200}

The larger RV variations from LAMOST observations, $\approx400$ km/s in about 1 hour,
motivated the interest of follow up observations.
In 2021, 6 exposures from 2 successive nights are obtained,
with the Double Spectrograph (DBSP) instrument on the Palomar 200-inch Hale Telescope (P200) \footnote{http://info.bao.ac.cn/tap/}.
The dichroic filter was set to $D\-55$. 
For the blue side, 
the grating was set to 1200 lines/mm, 5000 {\AA} blaze,
and the angle was set to $35^\circ52^\prime$
(center wavelength: 4500 {\AA}).
For the red side,
the grating was set to 1200 lines/mm, 7100 {\AA} blaze,
and the angle was set to $42^\circ02^\prime$
(center wavelength: 6600 {\AA}).
Data are processed with IRAF \footnote{http://iraf.noao.edu/specatlas/} and related information is shown in Table \ref{tab.sp_obs}.
These spectra cover wavelength range of 3900 to 5200 {\AA} in the blue band 
and 6250 to 7400 {\AA} in the red band, with S/N between 50 and 90.
The slit is set to be 2.0 arc seconds on Feb. 5 2021, and 1.5 arc seconds on Feb. 6 2021,
based on the weather condition of the site on these nights.
The resolution,
according to the P200 website and the slit configuration,
is $\sim2100$ in the blue
and $\sim3200$ in the red.
Since we aimed to get more RV measurements, 
no flux standard star was observed on these nights.

\begin{table*}
\caption{Spectroscopic observations and estimated parameters of J0338.
Note the DATE column represents the time zone observation night,
and the BJD represents the barycentric time at the middle of exposure.
The radial velocity has been corrected to the barycenter,
and the phase smearing effect is also corrected.
\label{tab.sp_obs}}
\setlength{\tabcolsep}{4.0pt}
\begin{center}
\begin{tabular}{ccccccccc}
\hline\noalign{\smallskip}
DATE & BJD &   RV  & Phase & $T_{\rm eff}$ & log $g$ & S/N & EXPTIME & Tel.$/$Ins. \\
     & day & km/s  &       & K             & dex     &     & s       &             \\
\hline\noalign{\smallskip} 
20150112 & 2457034.9594 & -312.62$\pm$3.14 & 0.31 & 22550$\pm$152 & 5.61$\pm$0.02 & 48 & 1500 & LAMOST/LRS \\ 
20150112 & 2457034.9789 & -109.90$\pm$3.78 & 0.46 & 22190$\pm$159 & 5.63$\pm$0.02 & 43 & 1500 & LAMOST/LRS \\ 
20150112 & 2457034.9983 & 133.16$\pm$3.62 & 0.62 & 22261$\pm$175 & 5.55$\pm$0.03 & 40 & 1500 & LAMOST/LRS \\ 
20150113 & 2457035.9649 & -285.07$\pm$4.55 & 0.33 & 22047$\pm$192 & 5.44$\pm$0.03 & 36 & 1500 & LAMOST/LRS \\ 
20150113 & 2457035.9850 & -50.13$\pm$4.52 & 0.49 & 22194$\pm$200 & 5.49$\pm$0.03 & 33 & 1500 & LAMOST/LRS \\ 
20150113 & 2457036.0052 & 193.36$\pm$4.68 & 0.65 & 22240$\pm$200 & 5.53$\pm$0.03 & 31 & 1500 & LAMOST/LRS \\ 
20160110 & 2457398.0173 & -38.06$\pm$4.54 & 0.51 & 21436$\pm$164 & 5.54$\pm$0.02 & 38 & 1500 & LAMOST/LRS \\ 
20160110 & 2457398.0367 & 214.02$\pm$5.04 & 0.66 & 21664$\pm$200 & 5.46$\pm$0.03 & 30 & 1500 & LAMOST/LRS \\ 
20160110 & 2457398.0680 & 94.59$\pm$3.76 & 0.91 & 21610$\pm$157 & 5.57$\pm$0.02 & 41 & 1500 & LAMOST/LRS \\ 
20220925 & 2459848.3008 & 241.51$\pm$6.63 & 0.78 & 22040$\pm$200 & 5.48$\pm$0.04 & 25 & 600 & LAMOST/LRS \\ 
20220925 & 2459848.3160 & 135.15$\pm$6.08 & 0.91 & 22865$\pm$200 & 5.68$\pm$0.04 & 27 & 600 & LAMOST/LRS \\ 
20220925 & 2459848.3320 & -91.98$\pm$7.93 & 0.03 & 22821$\pm$200 & 5.61$\pm$0.05 & 22 & 600 & LAMOST/LRS \\ 
20220925 & 2459848.3605 & -333.22$\pm$5.01 & 0.26 & 22886$\pm$200 & 5.61$\pm$0.03 & 33 & 600 & LAMOST/LRS \\ 
20220925 & 2459848.3688 & -299.17$\pm$5.09 & 0.33 & 22731$\pm$200 & 5.61$\pm$0.03 & 33 & 600 & LAMOST/LRS \\ 
20220926 & 2459849.2866 & 175.43$\pm$13.24 & 0.65 & 22632$\pm$200 & 5.62$\pm$0.08 & 14 & 300 & LAMOST/LRS \\ 
20220926 & 2459849.2942 & 234.47$\pm$9.49 & 0.71 & 21960$\pm$200 & 5.56$\pm$0.06 & 17 & 300 & LAMOST/LRS \\ 
20220926 & 2459849.2997 & 228.74$\pm$9.57 & 0.76 & 22178$\pm$200 & 5.67$\pm$0.05 & 20 & 300 & LAMOST/LRS \\ 
20220926 & 2459849.3053 & 230.03$\pm$7.37 & 0.80 & 22810$\pm$200 & 5.60$\pm$0.05 & 22 & 300 & LAMOST/LRS \\ 
20220926 & 2459849.3109 & 180.43$\pm$7.12 & 0.85 & 22787$\pm$200 & 5.75$\pm$0.05 & 24 & 300 & LAMOST/LRS \\ 
20220926 & 2459849.3213 & 86.77$\pm$10.65 & 0.93 & 24490$\pm$200 & 5.83$\pm$0.06 & 19 & 300 & LAMOST/LRS \\ 
20220926 & 2459849.3275 & -5.80$\pm$10.41 & 0.98 & 21694$\pm$200 & 5.40$\pm$0.06 & 19 & 300 & LAMOST/LRS \\ 
20220926 & 2459849.3386 & -199.85$\pm$13.44 & 0.07 & 23683$\pm$200 & 5.75$\pm$0.07 & 16 & 300 & LAMOST/LRS \\ 
20220926 & 2459849.3449 & -234.37$\pm$23.96 & 0.12 & 21227$\pm$200 & 5.41$\pm$0.07 & 15 & 300 & LAMOST/LRS \\ 
20220926 & 2459849.3557 & -315.80$\pm$6.81 & 0.20 & 22879$\pm$200 & 5.69$\pm$0.04 & 25 & 600 & LAMOST/LRS \\ 
20220929 & 2459852.2684 & -156.93$\pm$4.38 & 0.45 & 23358$\pm$200 & 5.74$\pm$0.03 & 40 & 600 & LAMOST/LRS \\ 
20220929 & 2459852.2805 & 19.76$\pm$20.43 & 0.54 & 21847$\pm$200 & 5.47$\pm$0.08 & 15 & 300 & LAMOST/LRS \\ 
20220929 & 2459852.2923 & 155.32$\pm$7.28 & 0.64 & 22671$\pm$200 & 5.60$\pm$0.05 & 24 & 300 & LAMOST/LRS \\ 
20220929 & 2459852.3124 & 240.25$\pm$8.30 & 0.80 & 22911$\pm$200 & 5.68$\pm$0.06 & 21 & 300 & LAMOST/LRS \\ 
20220929 & 2459852.3229 & 179.83$\pm$5.67 & 0.88 & 23048$\pm$200 & 5.63$\pm$0.04 & 30 & 300 & LAMOST/LRS \\ 
20220929 & 2459852.3333 & 27.01$\pm$10.20 & 0.96 & 23800$\pm$200 & 5.71$\pm$0.07 & 17 & 300 & LAMOST/LRS \\ 
20221006 & 2459859.3189 & 236.85$\pm$3.91 & 0.71 & 22601$\pm$184 & 5.57$\pm$0.03 & 43 & 600 & LAMOST/LRS \\ 
20210205 & 2459250.6901 & 224.31$\pm$2.70 & 0.85 & 23217$\pm$177 & 5.66$\pm$0.03 & 80 & 900 & P200/DBSP \\ 
20210205 & 2459250.7013 & 91.87$\pm$2.90 & 0.94 & 21993$\pm$158 & 5.50$\pm$0.02 & 69 & 600 & P200/DBSP \\ 
20210205 & 2459250.7085 & -5.52$\pm$2.90 & 1.00 & 22395$\pm$161 & 5.57$\pm$0.03 & 66 & 600 & P200/DBSP \\ 
20210206 & 2459251.6538 & 33.18$\pm$5.00 & 0.54 & 23247$\pm$200 & 5.68$\pm$0.05 & 65 & 600 & P200/DBSP \\ 
20210206 & 2459251.6633 & 149.56$\pm$5.00 & 0.61 & 22795$\pm$200 & 5.60$\pm$0.05 & 65 & 600 & P200/DBSP \\ 
20210206 & 2459251.6691 & 202.31$\pm$5.00 & 0.66 & 22525$\pm$200 & 5.57$\pm$0.06 & 44 & 300 & P200/DBSP \\ 
\noalign{\smallskip}\hline
\end{tabular}
\end{center}
\smallskip
\end{table*}

\subsection{Light curves}
\label{sect:lcs}

Since Dec. 2014, ASAS-SN \footnote{https://asas-sn.osu.edu/},
the \textbf{A}ll-\textbf{S}ky \textbf{A}utomated \textbf{S}urvey for \textbf{S}uper\textbf{N}ovae project,
has been monitoring the sky region of J0338 in V and g filters.
With the 24 globally distributed telescopes, 
ASAS-SN has the ability to monitor the entire sky every night down to about 18 magnitude.
Until recently, J0338 has been observed over 1200 times in g band and about 700 times in V band,
with the data available on the website
\footnote{https://asas-sn.osu.edu/sky-patrol/coordinate/603e442e-f770-4c1a-8307-b6bb485fbd75}. 
The g band data has a median value of 15.05 mag and a $\sigma$ of 0.08 mag,
while the V band data has a median value of 15.12 mag and a $\sigma$ of 0.09 mag.
No clear trend and periodic variations are found.

\begin{figure}
\center
\includegraphics[width=0.48\textwidth]{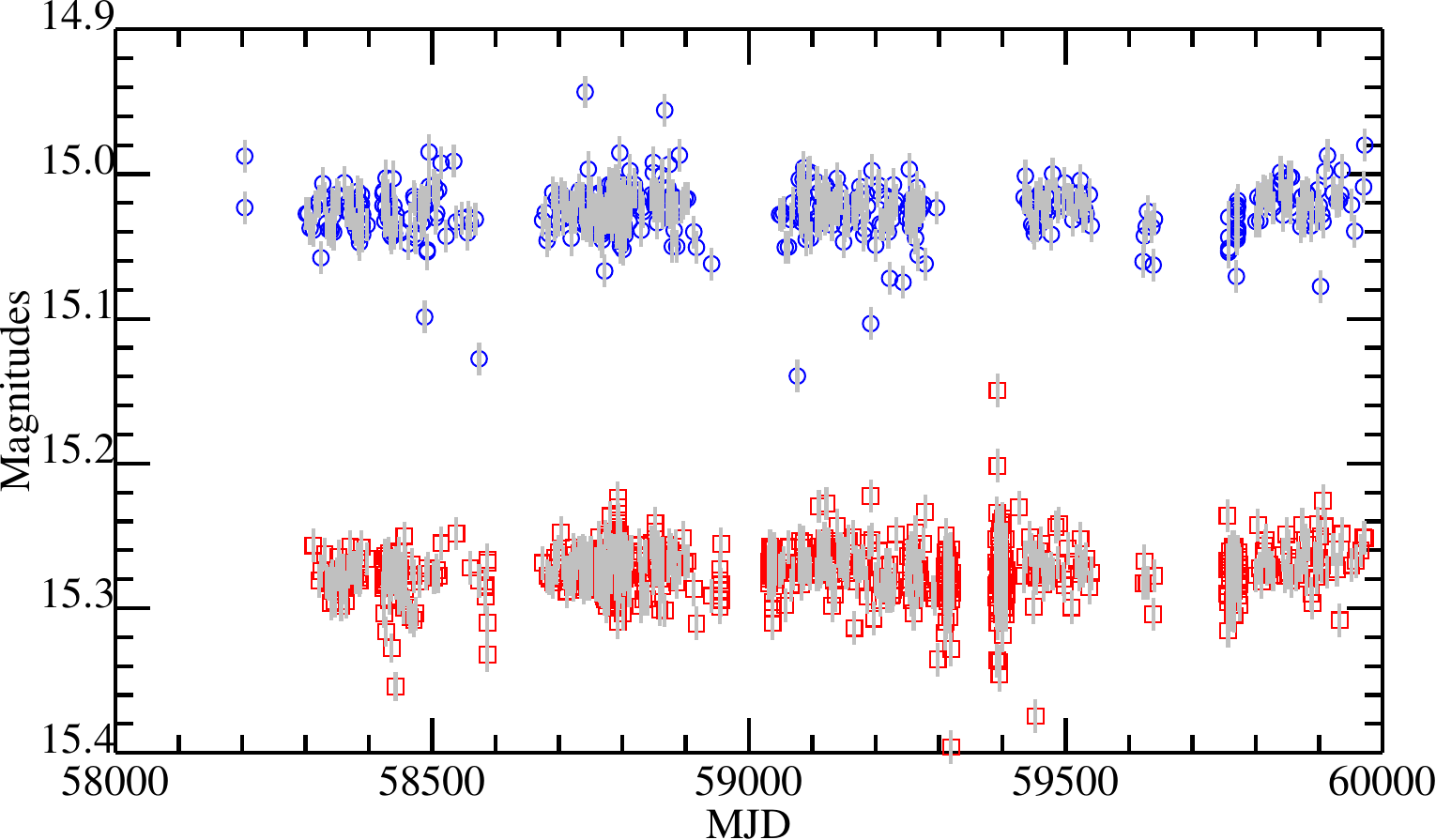}
\caption{
The ZTF g and r band light curves of J0338.
The blue circles represent g band observations, while the red boxes are for r band.
}
\label{lcztf}
\end{figure}

From Aug. 2018, ZTF \footnote{https://www.ztf.caltech.edu/},
\textbf{the} \textbf{Z}wicky \textbf{T}ransient \textbf{F}acility,
has also been monitoring the sky field of J0338, in g and r bands.
With the 48 inch Samuel Oschin telescope (P48) and the 47 square degree field of view, 
ZTF can survey the sky at 3750 square degrees per hour, down to about 20.4 magnitude.
Until recently, J0338 has been observed about 400 times in g band and over 700 times in r band.
The g band data has a median value of 15.03 mag and a $\sigma$ of 0.020 mag,
while the r band has a median value of 15.27 mag and a $\sigma$ of 0.036 mag.
The light curves are shown in Figure \ref{lcztf}.

\begin{figure}
\center
\includegraphics[width=0.48\textwidth]{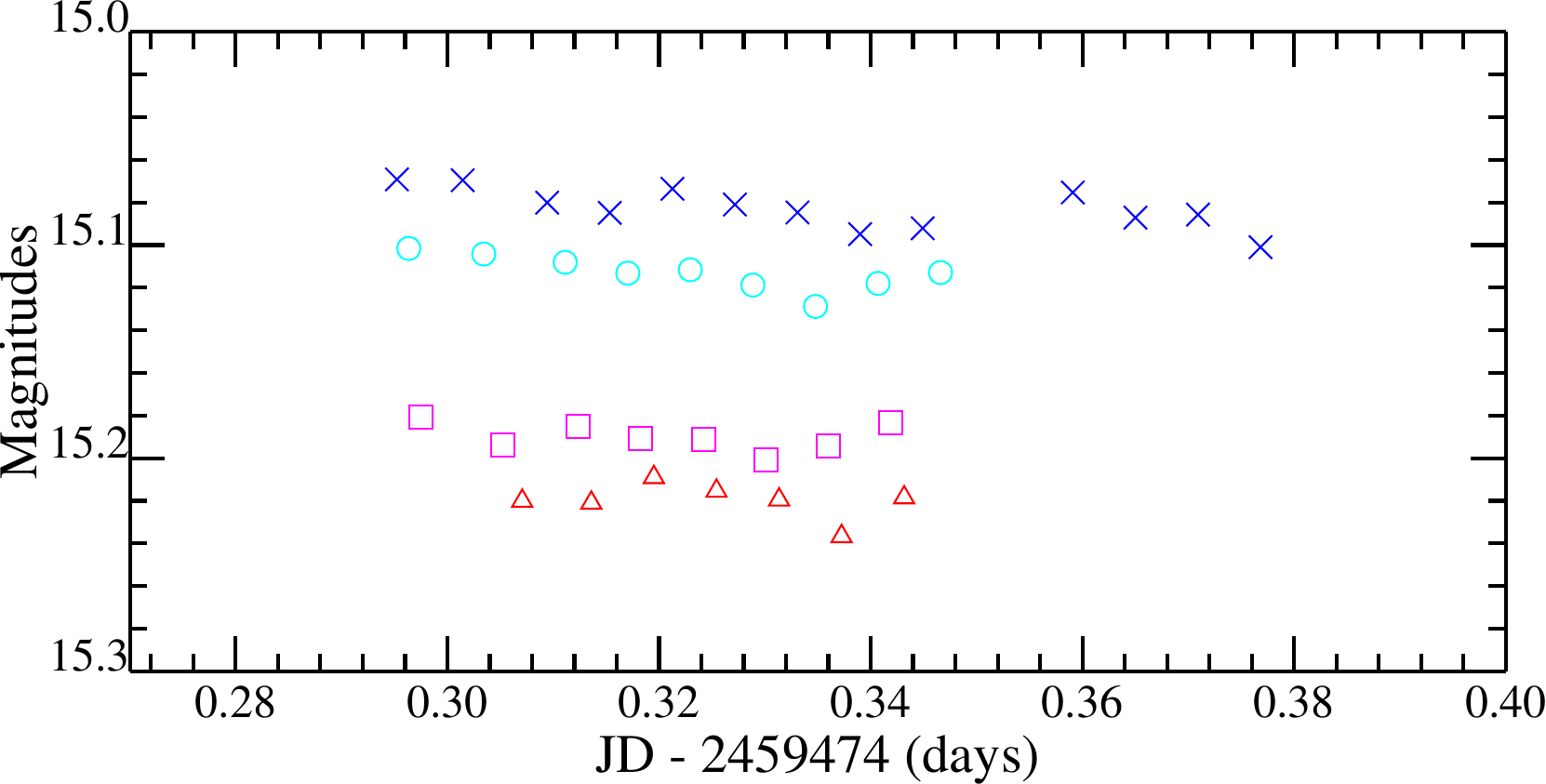}
\caption{
The 85cm telescope observations of J0338 on Sep 16 2021.
The B V R and I bands are denoted as blue crosses, cyan circles, magenta boxes and red triangles, respectively.
}
\label{lc85cm}
\end{figure}

On Sep. 16 2021, we have used the 85cm telescope at Xinglong Observatory to do photometry observation with BVRI filters.
Unfortunately, the weather was cloudy, and only a dozen measurements were obtained.
After bias subtraction and flat correction, the images are processed with Sextrator \footnote{https://www.astromatic.net/software/sextractor/},
The magnitudes are calibrated to B/V magnitudes from UCAC4 \citep{ucac4} and RI magnitudes from USNO-B1.0 \citep{usnob1.0}
($B=15.083\pm0.011, V=15.113\pm0.009, R=15.19, I=15.22$).
The result observations are shown in Figure \ref{lc85cm}.
The $\sigma$s of B V R and I bands returned by Sextrator, are 0.009, 0.008, 0.006 and 0.008, respectively.
Due to the stochastic cloud condition, 
the actual uncertainties of these points are difficult to be estimated.

\section{Data Analysis}
\label{sect:result}

\subsection{Distance and Extinction}
\label{sect:photo}

The multi-band photometric data are summaried in Table \ref{tab.mags}.
The Gaia early release DR3 \citep{2021Gaia} has reported the parameters of J0338:
G band magnitude $Gmag=15.088968\pm0.002838$, parallax $Plx=1.6469\pm0.0318$ mas,
color $BR-RP=-0.0614$ and $RUWE=0.949$.
The parallax zero point, according to \citet{gaiaedr3plxzero} is $-0.0306$ mas.
Thus the corrected parallax is $Plx=1.6775\pm0.0318$ mas, 
corresponding to a distance of $Dkpc=0.596\pm0.011$ kpc.
The distance modules (DM) can be calculated as $8.876\pm0.040$.
%
With this distance, 
\citet{GreenDust2019} provides a 3D extinction of $E(B-V)=0.16\pm0.02$.
A previous version of 3D extinction \citep[][]{GreenDust2017} provides a value of $0.12\pm0.02$.
The 2D dust maps of \citet[][]{Schlegel1998} and \citet[][]{Planck2016Dustmap}
provide a value of $\sim0.21$ and $\sim0.24$, respectively.
J0338 is $\sim600$ pc away so the extinction should be smaller than the 2D estimations.
As will be discussed in the following sections,
the visible component has a low filling factor and the companion is a compact white dwarf,
we expect no significant circumbinary dust extinction.
Thus an extinction of $E(B-V)=0.16$ is an appropriate estimation. We use a larger uncertainty of $0.06$ mag and will verify it in (Section \ref{sect:sed}).


\begin{table*}
\caption{Multi-band photometry for J0338. \label{tab.mags}}
\setlength{\tabcolsep}{4.5pt}
\begin{center}
\begin{threeparttable}
\begin{tabular}{cccc}
\hline\noalign{\smallskip}
Filter & Magnitude & Reference \\
\hline\noalign{\smallskip}
GALEX FUV & 15.1771$\pm$0.0145\tnote{1} & \citet[][]{Bianchi2017galex} \\
GALEX NUV & 15.4488$\pm$0.015\tnote{1}  & \citet[][]{Bianchi2017galex} \\
SDSS u & 15.152$\pm$0.004 & \citet[][]{2020ApJS..249....3A} \\
SDSS g & 14.961$\pm$0.004 & \citet[][]{2020ApJS..249....3A} \\
SDSS r & 15.212$\pm$0.004 & \citet[][]{2020ApJS..249....3A} \\
SDSS i & 15.436$\pm$0.004 & \citet[][]{2020ApJS..249....3A} \\
SDSS z & 15.668$\pm$0.006 & \citet[][]{2020ApJS..249....3A} \\
PS1 g & 15.0485$\pm$0      & \citet[][]{2016arXiv161205560C} \\
PS1 r & 15.2401$\pm$0.0033 & \citet[][]{2016arXiv161205560C} \\
PS1 i & 15.4619$\pm$0.0018 & \citet[][]{2016arXiv161205560C} \\
PS1 z & 15.6578$\pm$0.0026 & \citet[][]{2016arXiv161205560C} \\
PS1 y & 15.7622$\pm$0.0048 & \citet[][]{2016arXiv161205560C} \\
GaiaDR2 G  & 15.1104$\pm$0.0006 & \citet[][]{2018gaia} \\
GaiaDR2 BP & 15.0390$\pm$0.0049 & \citet[][]{2018gaia} \\
GaiaDR2 RP & 15.1303$\pm$0.0019 & \citet[][]{2018gaia} \\
GaiaEDR3 G  & 15.0890$\pm$0.0028 & \citet[]{2021Gaia} \\
GaiaEDR3 BP & 15.0634$\pm$0.0031 & \citet[]{2021Gaia} \\
GaiaEDR3 RP & 15.1249$\pm$0.0042 & \citet[]{2021Gaia} \\
2MASS J	& 15.322$\pm$0.043\tnote{2} & \citet[][]{Cutri2003_2mass} \\
2MASS H & 15.375$\pm$0.094\tnote{2} & \citet[][]{Cutri2003_2mass} \\
2MASS K & 15.335$\pm$0.17\tnote{2}  & \citet[][]{Cutri2003_2mass} \\
WISE W1 & 15.389$\pm$0.040\tnote{3} & \citet[][]{Cutri2014_wise} \\
WISE W2 & 15.784$\pm$0.154\tnote{3} & \citet[][]{Cutri2014_wise} \\
WISE W3 & 12.496\tnote{3} & \citet[][]{Cutri2014_wise} \\
WISE W4 & 8.762\tnote{3} & \citet[][]{Cutri2014_wise} \\ 
\noalign{\smallskip}\hline
\end{tabular}
\begin{tablenotes}
\footnotesize
\item[1] All the artifact and extraction flags for GALEX FUV and NUV bands are negative.
\item[2] The Q/R/B/C flags for 2MASS are {\bf AAC/222/111/000}. 
\item[3] The quality flag for WISE is {\bf ABUU}, meaning only W1 is reliable.
\end{tablenotes}
\end{threeparttable}
\end{center}
\end{table*}

\subsection{Spectral Parameters}
\label{sect:sp}

Spectral RV measurements are carried out by minimizing $\chi^2$ values of full spectra template fitting, 
and the result is corrected to the barycenter \citep{Bai2021}.
It is necessary to mention that as shown in Table 4 of \citet{Bai2021}, 
spectra of hot stars show only broad absorption lines may have large system errors 
($\approx9.77$ km/s for different days, $\approx4.97$ km/s for the same day, for A-type stars).
Using the system velocity from Section \ref{sect:orbit} and method used in \citet[][]{Yuan2023},
the phase smearing correction up to several km/s are applied.
The corrected RVs, and their fitting errors, are shown in Table \ref{tab.sp_obs}.

\begin{figure}
\center
\includegraphics[width=0.48\textwidth]{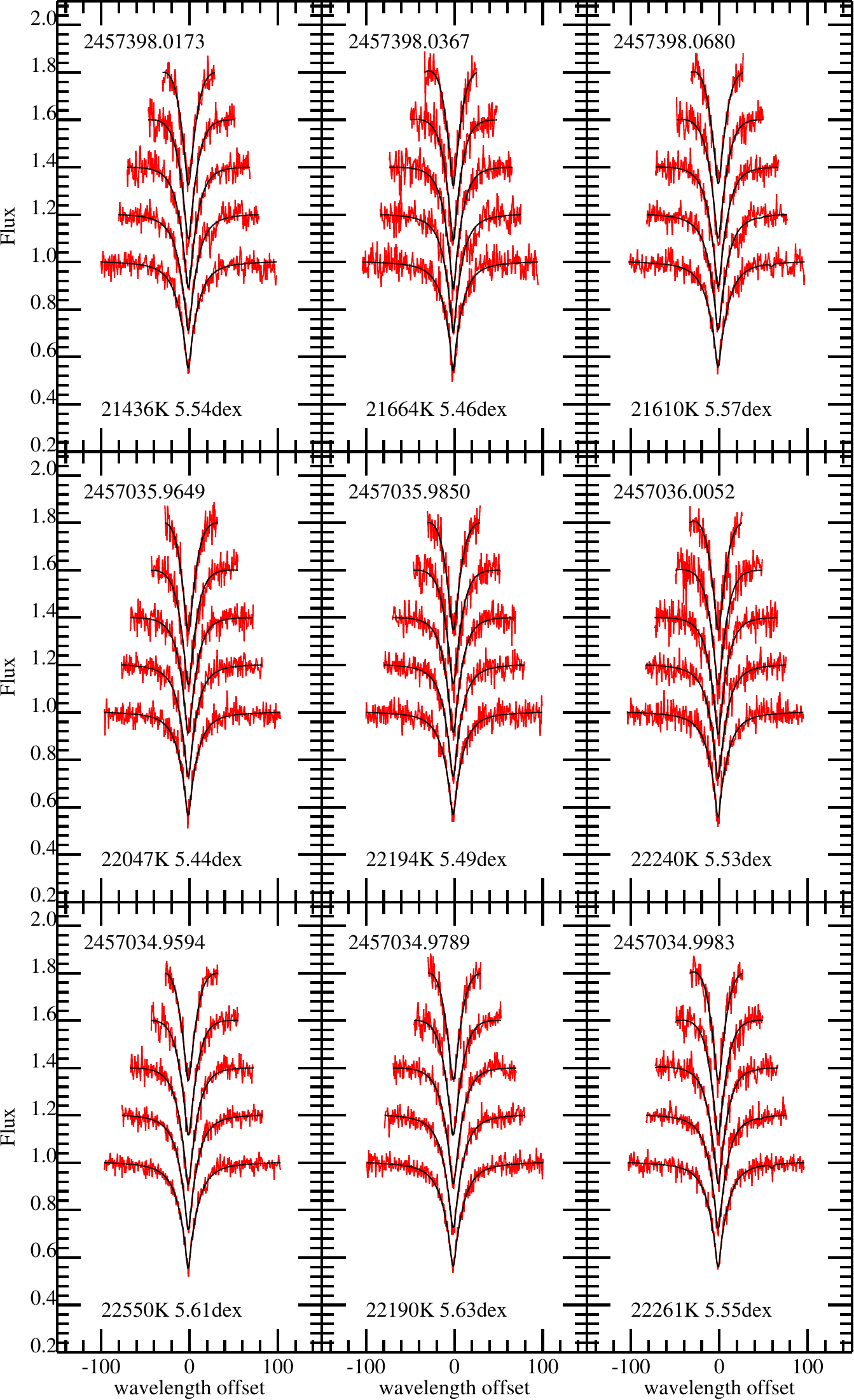}
\caption{
Model fits to the first 9 single exposure spectra of J0338. 
The red lines denote observations and the black ones denote model spectra.
Note wavelength offset represents offsets in unit of {\AA} from the central H lines.
The barycentric Julian days and best fit parameters are shown in the figure.
The lines range from $H\beta$(bottom) to $H8$(top) and adjacent lines are vertically shifted by 0.2 for clarity.
}
\label{spfit}
\end{figure}

All these spectra are dominated by broad H Balmer lines,
without any other clear features, like emission, Helium absorption and metal lines,
as shown in Figure \ref{sp_blue}.
The non-LTE spectral template grid is built from {\bf TLUSTY} and {\bf SYNSPEC} code \citep[][]{Hubeny1988TLUSTY, 2014ASPC..481...95N},
and matched to these single exposure spectra \citep{Luo2016sdB, Lei2018sdB}.
The models are interpolated to specified parameters, Gaussian convolved to the instrument resolution, 
and multiplied with a polynomial curve to fit the observed spectra.
Note the spectral resolution is smaller on the blue band,
and is different on different days, from $\sim1000$ to $\sim2000$
based on arc line fitting results at different wavelengths.
Only the five spectral regions, $H_\beta, H_\gamma, H_\delta$, $H_\epsilon$ and $H_\zeta$, are considered \citep{wdparameter}.
For P200 spectra, the $H_\zeta$ is also ignored as the quality is low in that region.
The Downhill algorithm is adopted to find the best parameters.
The results are demonstrated in Table \ref{tab.sp_obs} and Figure \ref{spfit}.

Preliminary tests show that 
the atmosphere parameters from P200 spectra without flux calibration have larger surface gravity by $\sim0.3$ dex,
which will seriously affect the estimation of the visible component's mass and nature.
In order to better estimate the parameters,
we have made relative flux calibration using an F-type star observed on the same night.
The response curve is calculated by dividing that observed spectra with theoretic spectra.
Then the curve is median smoothed with a window of $\sim100$ {\AA} before applied to correct the J0338.
We have also ignored the $H_\zeta$ since the flux correction here has large uncertainty.
The refined estimations are in good agreement with the LAMOST results.

\begin{figure}
\center
\includegraphics[width=0.48\textwidth]{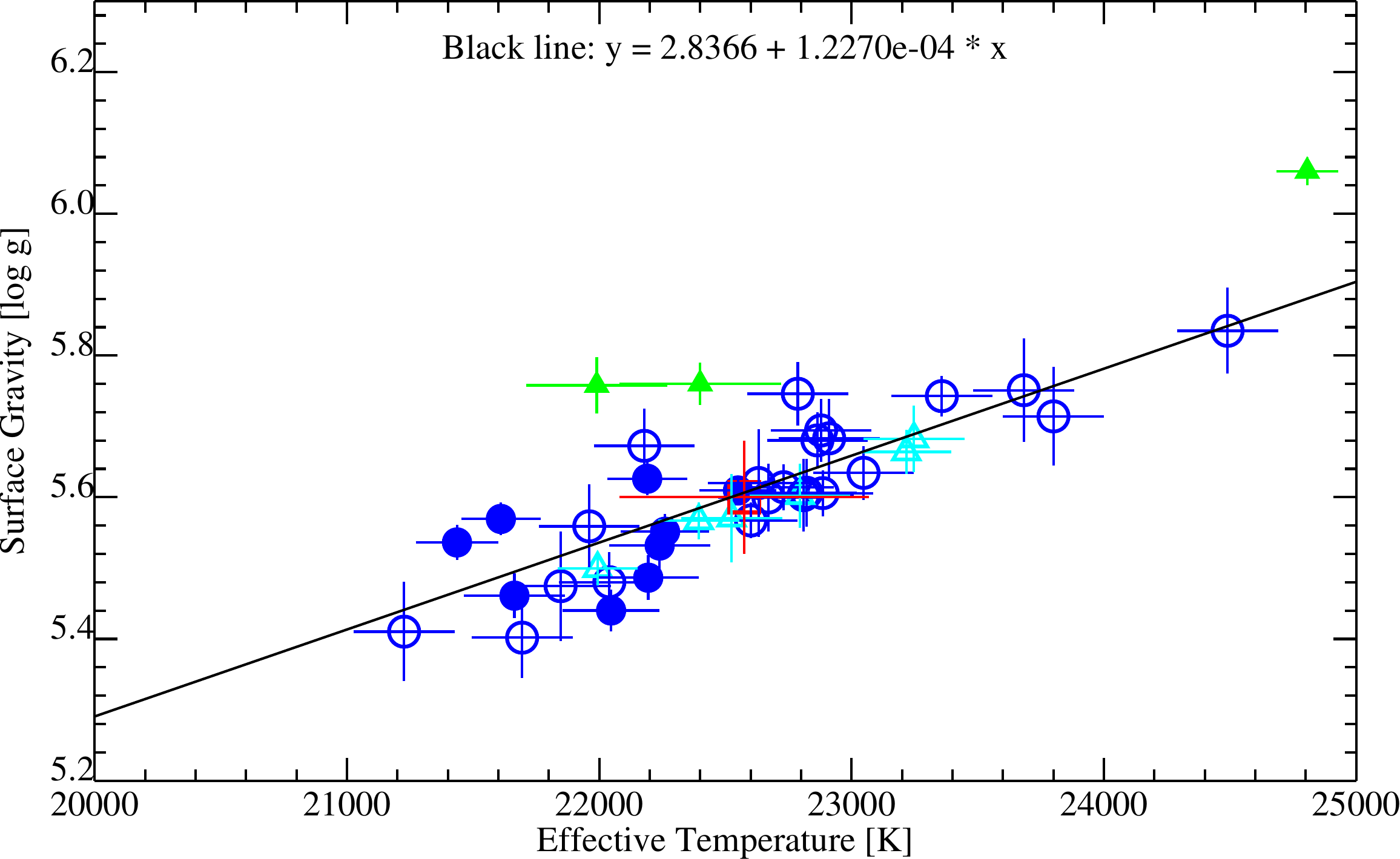}
\caption{
The correlation between effective temperature and surface gravity of J0338 
from single exposures.
The filled blue circles, open blue circles and cyan triangles,
denote the first 9 exposures, the other LAMOST exposures 
and the P200 exposures, respectively.
The S/N weighted parameters are shown as red rectangle.
The filled green triangles denote values from previous works.
The black oblique line denotes a linear fit between effective temperature and surface gravity.
See Section \ref{sect:sp} for details.
}
\label{fig:teff_logg}
\end{figure}

\begin{figure}
\center
\includegraphics[width=0.48\textwidth]{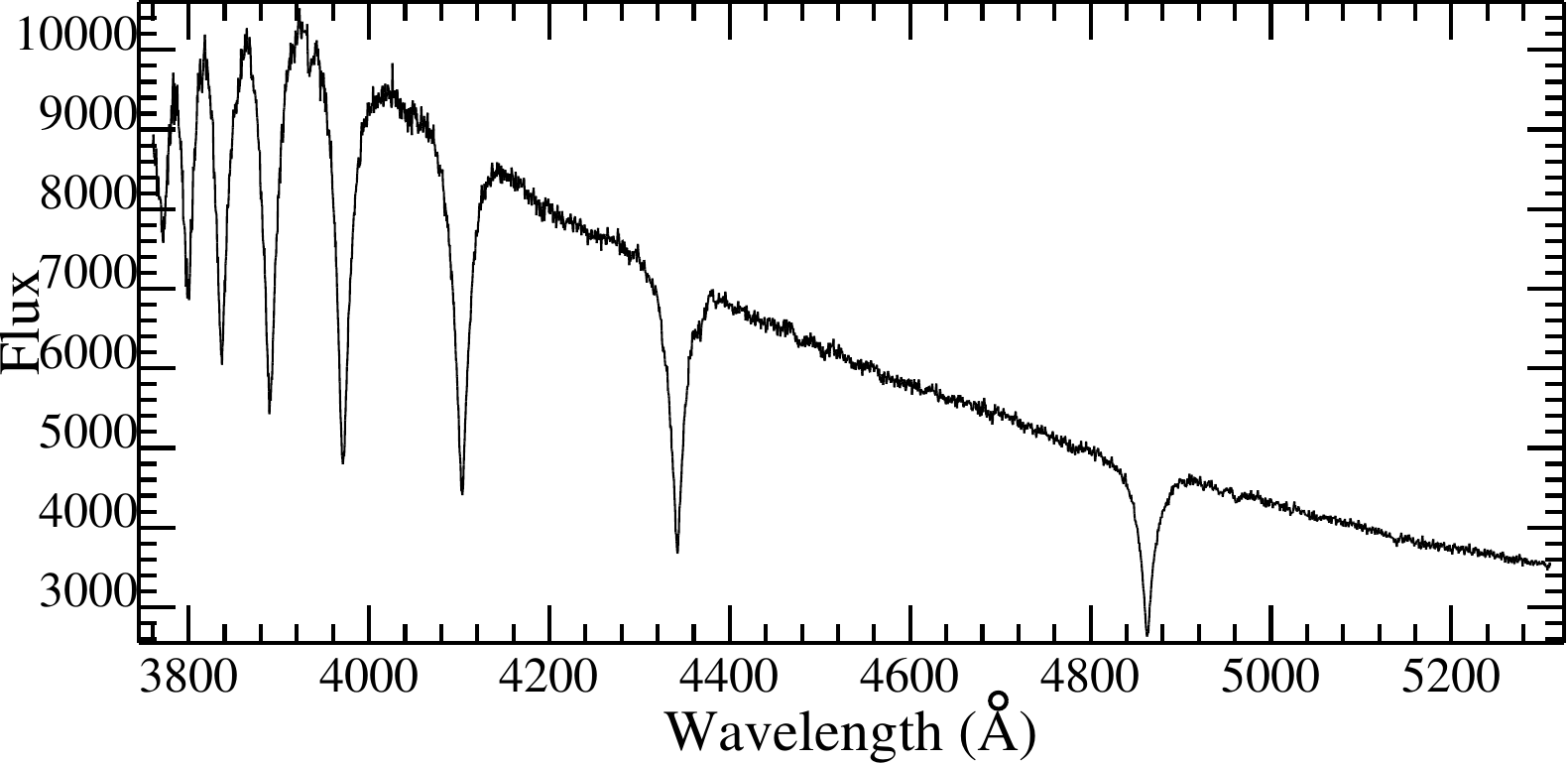}
\caption{
The combined blue band rest frame spectra of J0338. 
Note we only use the first nine single exposures, 
which have equal exposure duration of 1500s, and similar resolutions.}
\label{sp_blue}
\end{figure}

The measured atmosphere parameters show no correlation with orbital phase,
but there exists a positive correlation between temperature and gravity,
as show in Figure \ref{fig:teff_logg}.
This correlation might be intrinsic and reasonable,
since both lower temperature and larger gravity corresponds to broader H line profiles.
We have verified this effect with simulated spectra.
The uncertainties from the single exposure spectra fitting procedure are relatively small, 
but the deviations between different exposures are remarkable.
All single measurements are converted into Gaussian probability density distributions (PDF), 
normalized with S/N, and summed up into a combined PDF. 
Then the S/N weighted parameters can be calculated as $T_{\rm eff}=22575\pm494$ K, log $g=5.60\pm0.08$ dex.

To get higher S/N ratio spectra, 
we have also tried to combine single spectra in the rest frame.
As will be discussed in Section \ref{sect:vsini},
the difficult is the different exposure lengths and different RV variations at different orbital phases.
We tried to combine only the first 9 exposures, 
which have equal exposure duration of 1500s, and similar resolutions
(Figure \ref{sp_blue}).
The results are consistent with S/N weighted value above.
We have also noticed that,
\citet{Lei2018sdB} estimated the parameter of our target to be 
$T_{\rm eff}=22240\pm320$ K and log $g=5.76\pm0.03$ dex,
while \citet{Luo2021sdB} estimated the parameter to be 
$T_{\rm eff}=24806\pm123$ K and log $g=6.06\pm0.02$ dex.
In the work of \citet[][]{Wang2022ELM}, the estimated parameters are $T_{\rm eff}=21990\pm280$ K and $5.758\pm0.040$ dex.
These three measurements are also shown in Figure \ref{fig:teff_logg}.
The larger surface gravity may be caused by using a higher spectral resolution of $R\sim1800$, 
which is the average value for the whole spectra. 


\subsection{SED Fitting}
\label{sect:sed}

\begin{figure}
\center
\includegraphics[width=0.48\textwidth]{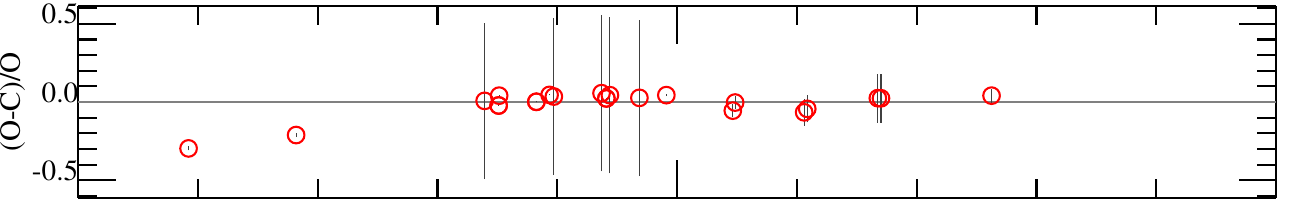}
\includegraphics[width=0.48\textwidth]{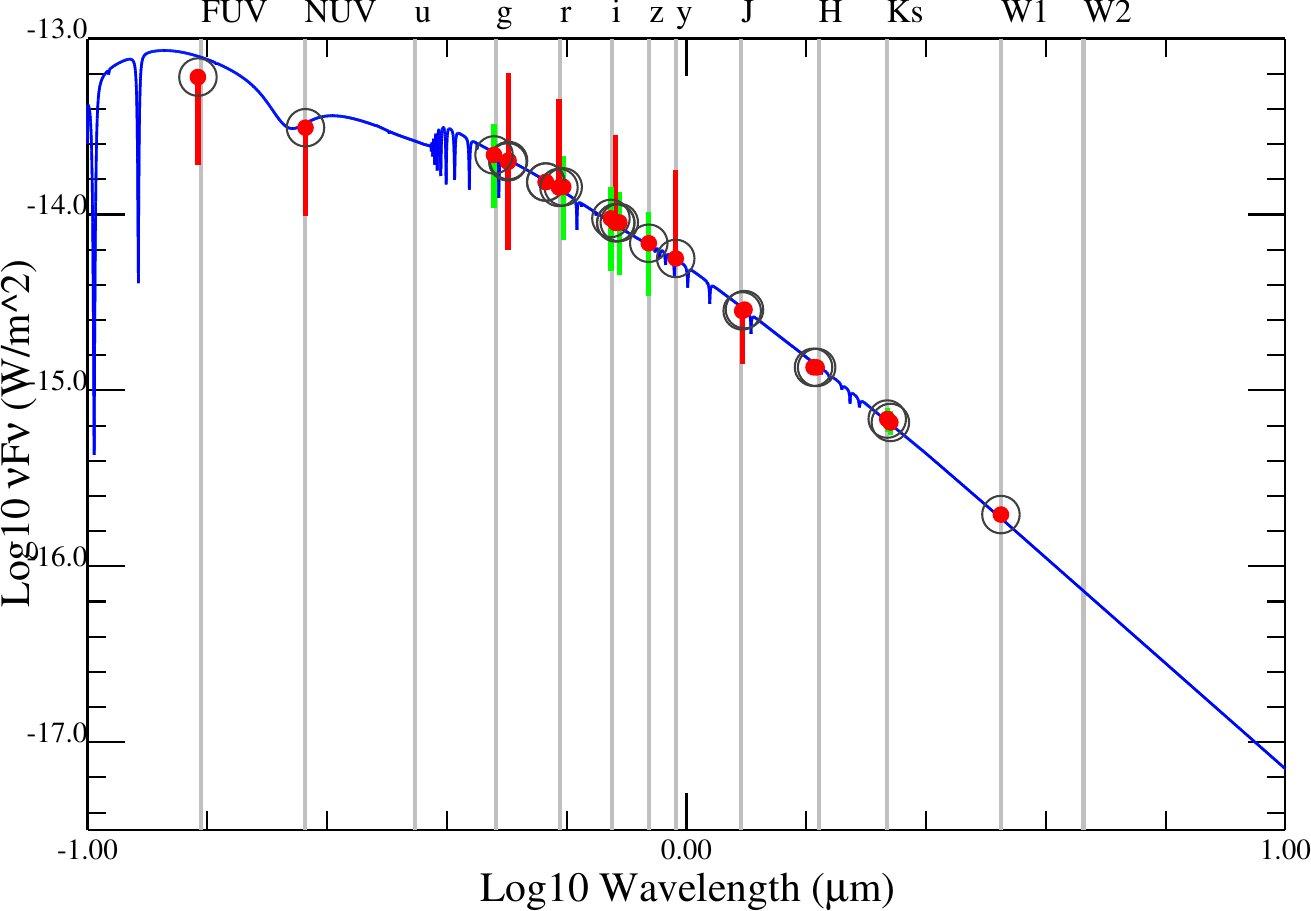}
\caption{
The SED fitting of J0338 using optical and near infrared photometric data.
The GALEX data are shown for reference and not used in the fitting since they deviate the spectral model by 20-30\%.
}
\label{sed}
\end{figure}

The Spectral Energy Distribution (SED) fitting, based on Optical and Infrared observations, 
is shown in Figure \ref{sed}.
The GALEX data includes 2 visits in FUV band and 1 visit in NUV band.
Trial tests show that the GALEX fluxes are lower than the fitting by 20-30\% percent.
The non-linear correlation for GALEX, especially for bright targets,
is discussed by many authors 
\citep[][]{Morrissey2007Galex, Camarota2014Galex, Wall2019Galex},
For J0338 the correction
is $\sim$5\% according to the observed magnitudes ($\sim15$),
thus the GALEX data points are not considered in the fitting process.
The model spectra is provided by the T$\ddot u$bingen NLTE Model-Atmosphere Package (TMAP) \citep[][]{TMAP}.
Four priors are considered:
$E(B-V)=0.16\pm0.06$ and parallax $=1.6775\pm0.0318$ mas from Section \ref{sect:photo};
$T_{\rm eff}$ and log $g$ from Section \ref{sect:sp}.
Under the Markov Chain Monte Carlo (MCMC) framework,
the SED fitting is carried out in away similar to \citep[][]{Yuan2023},
and the results are $Radius=0.121\pm0.003R_{\odot}$ and $E(B-V)=0.14\pm0.01$.
The posterior distributions of temperature and gravity show tiny difference with the priors.

\begin{figure}
\center
\includegraphics[width=0.48\textwidth]{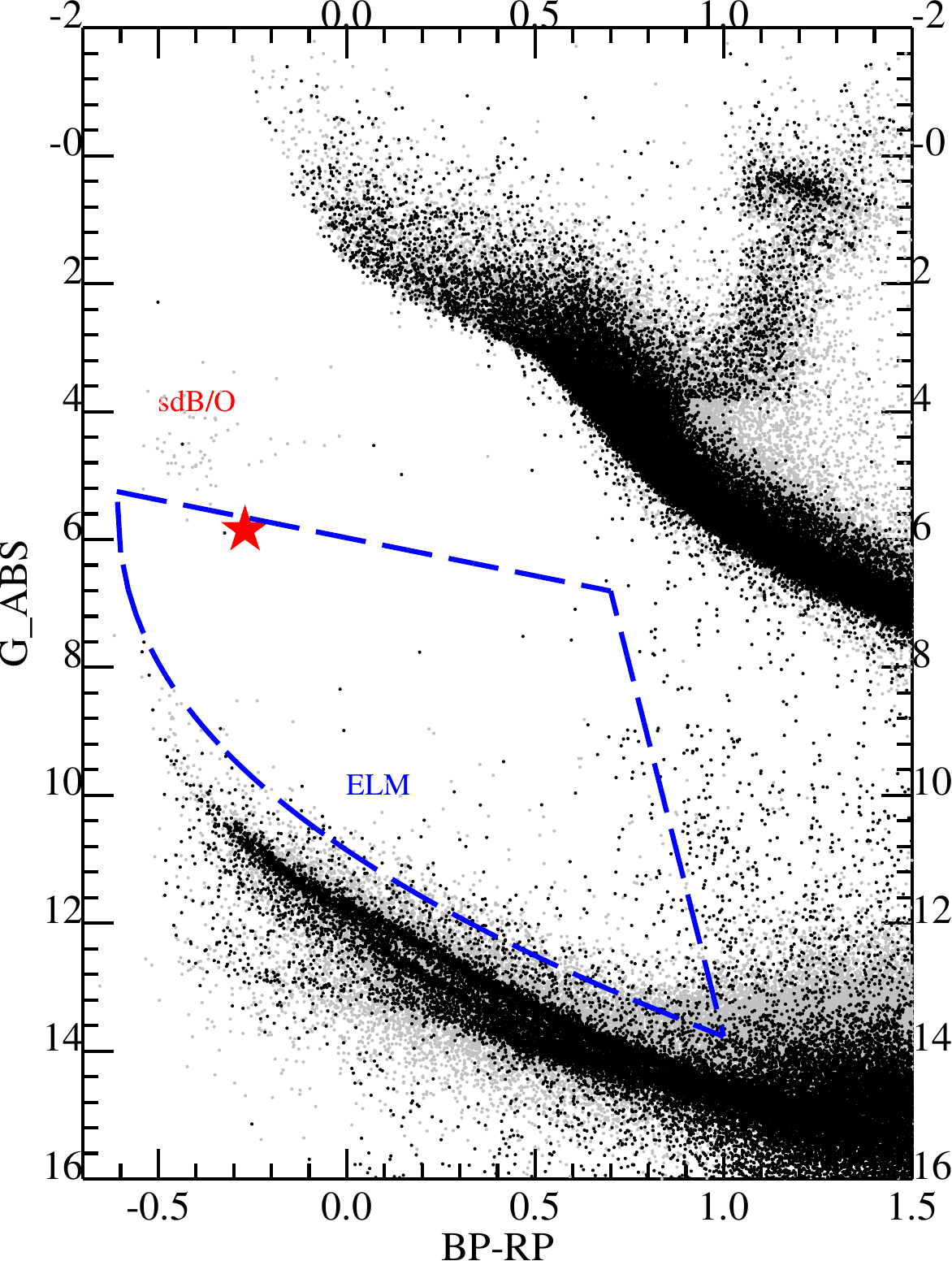}
\caption{
The position of J0338 (red star) in Gaia colour-magnitude diagram.
The absolute G band magnitude is calculated as $G_{abs}=G+5\log((plx+0.029)/1000)+5$,
and no reddening correction is applied.
The black and gray points are Gaia 100 and 300 pc samples, respectively,
using similar selection criteria of Figure 1 of \citet[][]{Pelisoli2019GaiaELM}.
The blue dashed line region is the ELM WD selection criteria of \citet[][]{Pelisoli2019GaiaELM},
which marginally classifies J0338 as an ELM WD.
}
\label{fig:cmd}
\end{figure}

As we would claim the unseen companion of J0338 to be a white dwarf (Section \ref{sect:orbit}),
the flux contribution of a white dwarf may be around 1\% or less, 
assuming the white dwarf is not very hot.
For example, a white dwarf of $40000$ K might contribute a flux contribution of $\sim10\%$,
in which case some features should be identified in the spectra.
With the result of SED fitting, the reddening corrected $G_{abs}$ and $G_{BP}-G_{RP}$
is $5.85\pm0.05$ and $-0.27\pm0.02$, respectively, using bootstrapping method.
In the Gaia $G_{abs}$ and $G_{BP}-G_{RP}$ color magnitude diagram (see Figure \ref{fig:cmd}), 
J0338 is located between the main sequence and the white dwarf cooling sequence,
and slightly below the hot subdwarf dense region.

\subsection{Mass of the visible component}
\label{sect:mass}

With the radius from SED fitting ($0.121\pm0.003R_{\odot}$) and the surface gravity from spectral fitting,
a bootstrapping procedure is carried out to calculate the mass, assuming a solar surface gravity of 4.4354, 
using formula
\begin{equation}
\label{eq.logg}
M/R^2 = 10^{({\rm log}\,g - 4.4354)}
\end{equation},
where $M$ and $R$ denotes mass and radius with respect to the Sun.
.
The resultant mass $M_{spec}$ is $0.22\pm0.05M_{\odot}$.
\citet[][]{Althaus2013ELM} presented a homogeneous grid of evolutionary sequence for He core white dwarfs.
By interpolating to the grid of \citet[][]{Althaus2013ELM} and taking into consideration the atmosphere parameter uncertainties, 
the mass, age and radius can be calculated, with bootstrapping method.
It is found that the two approaches appropriately agree with each other.
The resultant mass $M_{model}$ is $0.215\pm0.005$ $M_{\odot}$, 
and the corresponding radius, is $0.122\pm0.013$ $R_{\odot}$,
based on mass and gravity (Formula \ref{eq.logg}).
\citet[][]{Istrate2016ELMs} have also provided evolution and cooling tracks of ELMs, 
with effects of rotational mixing and element diffusion taken into consideration.
Using tracks of $Z=0.01$ and limiting points before H flash,
the related mass according the stellar parameters is $0.235\pm0.0055M_{\odot}$.
If the points after the H flash are considered, 
the tracks with masses down to $\sim0.18M_\odot$ can reach the measured stellar parameters of J0338. 
However, the parameters of J0338 is unlikely caused by the H flash due to the extremely short timescale ($1\;\rm Myr$, \citealt{Istrate2016ELMs}).
The final adopted parameters are summarized in Table \ref{tab.parameters}.

\subsection{Orbital parameters}
\label{sect:orbit}

\begin{figure}
\center
\includegraphics[width=0.48\textwidth]{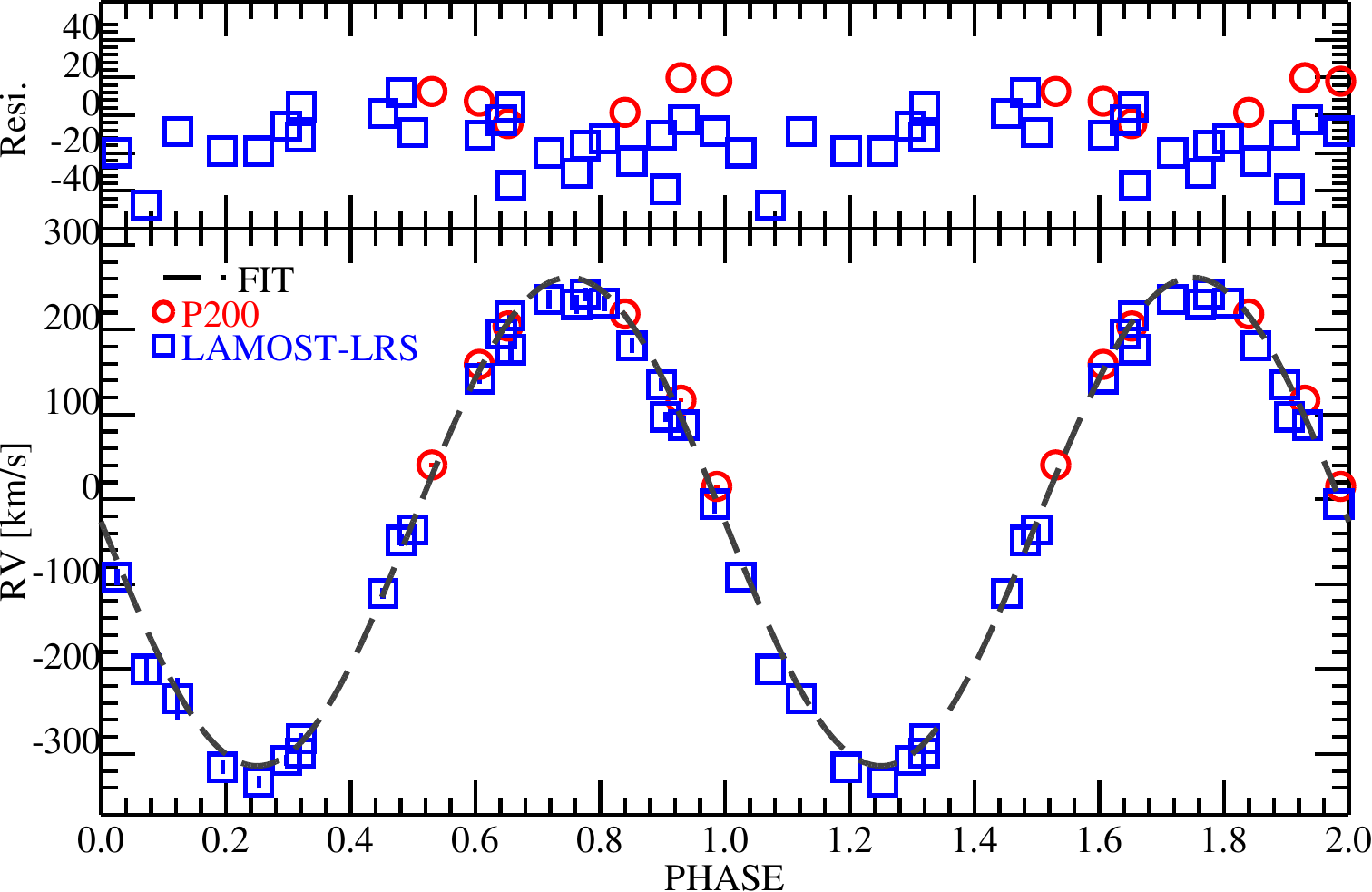}
\caption{
The folded velocity curve and fitting residuals of J0338.
The blue boxes denote the LAMOST/LRS observations and the cyan circles denote those from P200/DBSP.
The dashed line shows the best fit with $K1=289$ km/s and $V_0=-41$ km/s.
}
\label{rvfold}
\end{figure}

The RV data from both LAMOST and P200,
as shown in Table \ref{tab.sp_obs},
is used to solve the orbital parameters
within a large period range
by utilizing TheJoker \citep{2017ApJ...837...20P},
which is a well performed Monte Carlo sampler for sparse RV measurements.
The result peaks at period $\sim0.125$ days clearly.
Then we run TheJoker within a small range around 0.125days (from 0.1236 to 0.1261days),
with an expected sample size of 1024.
The returned orbital parameters are 
$p=0.1253132\pm0.0000001$ days,
$e=0.03\pm0.03$, 
and $K1\sim290$ km/s.
The sampled eccentricities show a distribution peak at 0.
We further carry out a sinusoidal curve fit ($ecc$ = 0),
using MCMC framework with a sample size of 5000.
We have repeated this step 16 times, 
and for each step we add random errors to the RV values according to their uncertainties.
The final sample contains 80,000 sets of parameters, 
and orbital parameters are estimated to be
$P=0.1253132\pm0.0000001$ days, $K1=289\pm4$ km/s, $V_{sys}=-41\pm3$ km/s,
and zero point for $\phi$ is $T_0=2457034.9211\pm0.0003$ days.
The folded velocity curve, together with the sinusoidal fit and residuals, are shown in Figure \ref{rvfold}.
Thus the orbital ephemeris of the system is,
\begin{equation}
T (\phi = 0) = 2457034.9211(3)BJD + 0.1253132(1) \times N,
\end{equation}.
Here $\bf BJD$ denotes barycenter Julian day with decimals.
As mentioned above, the phase 0 time denotes the moment where the visible component is at superior conjunction point.

\begin{figure}
\center
\includegraphics[width=0.48\textwidth]{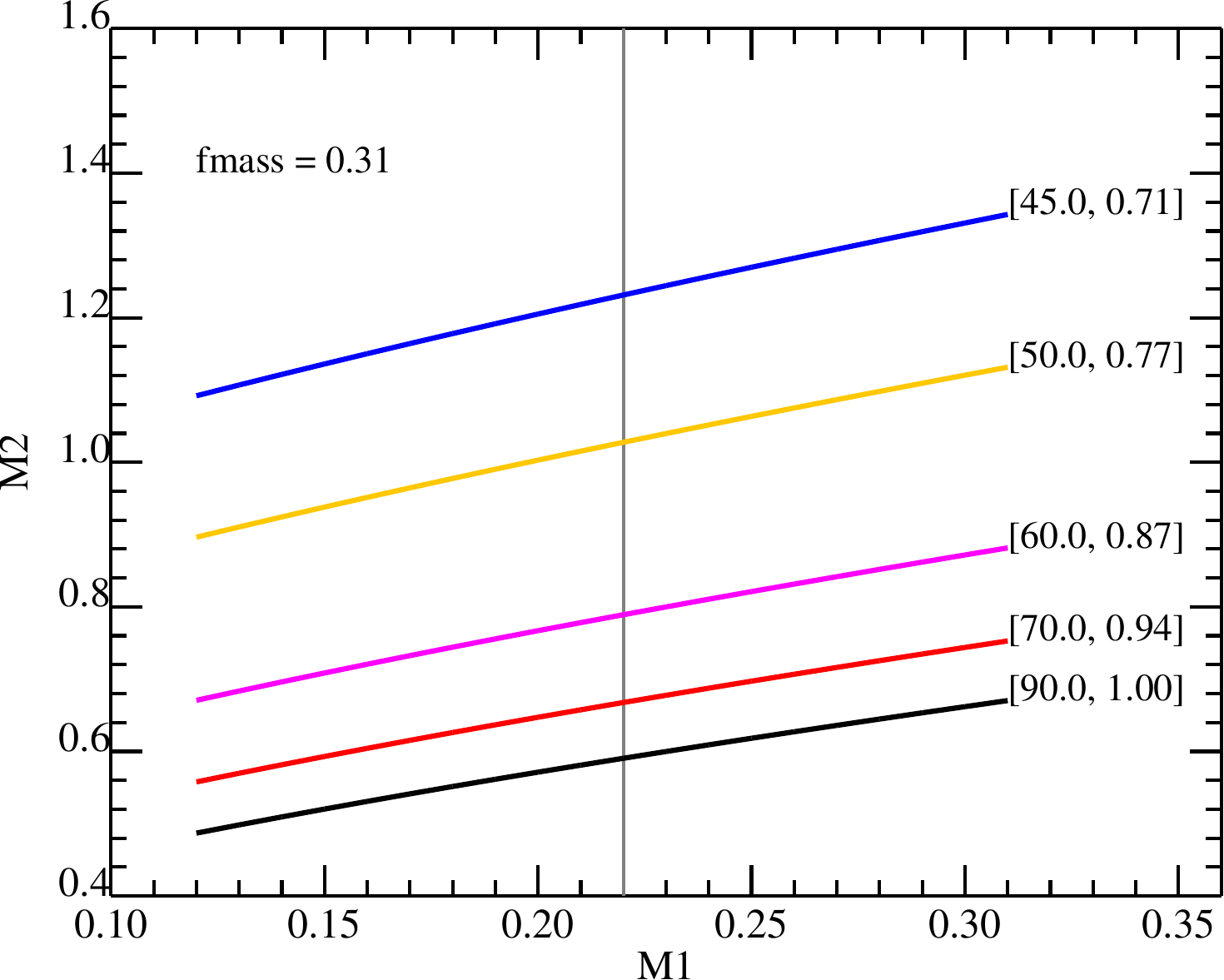}
\caption{
The m1 and m2 relationship based on mass function 0.31$M_{\odot}$.
The black, red, pink, yellow and blue lines represent different inclination angles, 
with $inc$ in degrees and $sini$ shown on the right side.
The vertical gray line denotes the mass estimation (Table \ref{tab.parameters}).
}
\label{k1_m1_m2_inc}
\end{figure}

Then the mass function of the secondary is calculated to be $\approx0.31$ $M_{\odot}$. 
The relationship between primary and secondary,
considering different orbital inclination angles, is demonstrated in Figure \ref{k1_m1_m2_inc}.
As described in Section \ref{sect:sp}, the mass of star 1 is $\approx0.22M_{\odot}$, 
thus the mass of the unseen companion is expected to be above 0.60$M_\odot$.
If a median inclination of 60.0$^{\circ}$ is considered, assuming a randomly orientated binary, the mass of the secondary could be $\approx0.79M_{\odot}$.
The two mass values well match the 2nd and 3rd peaks of the SDSS DA white dwarf mass distribution provided by \citet[][]{Kleinman2013whitedwarf}.

By assuming different inclination angles (from 15$^\circ$ to 90$^\circ$), 
the mass ratio ($q=m2/m1$, or the mass of secondary m2), 
orbital separation of the two components $SMA$,
Roche volume radius of star 1 $R_{Roche}$, 
and filling factor ($R1/R_{Roche}$) can all be calculated step by step,
utilizing binary mass function formula, 
the third Kepler's law and
the approximate analytical formula from \citet[][]{Eggleton1989},
\begin{equation}
\frac{R_{Roche}(2)}{SMA}=\frac{0.49q^{2/3}}{0.6q^{2/3}+\ln{1+q^{1/3}}}
\end{equation}
.
The calculation indicates that
as inclination angle increases,
the orbital separation ($\sim1.1R_{\odot}$) monotonously decreases from above 2 to around 1 $R_{\odot}$
while the relative Roche column radius $R_{Roche}/SMA$ increases monotonously,
yielding an approximately stable star 1 filling factor ($\approx43\%$),
which is relatively low for a close system.
The filling factor will be used to constrain the orbital inclination,
as shown in Section \ref{sect:lcperiod}.

\subsection{Galactic Membership}
\label{sect:membership}

The proper motion of J0338 in Gaia EDR3 is $pmra=-5.423\pm0.036$ mas/year and $pmdec=5.79\pm0.029$ mas/year.
Together with system velocity of $-41\pm3$ km/s from our orbit fitting above,
the 3D motion in LSR $(U,V,W)_{LSR}$ is ($44.4\pm2.6$km/s, $-4.5\pm1.3$ kms, $14.4\pm0.6$ kms),
indicating a thin disk object 
($P_{thin}\sim99\%$)
\citep[][]{2013ApJ...764...78R, Brown2020ELM}.



\subsection{Light curve period search}
\label{sect:lcperiod}

\begin{figure*}
\center
\includegraphics[width=0.96\textwidth]{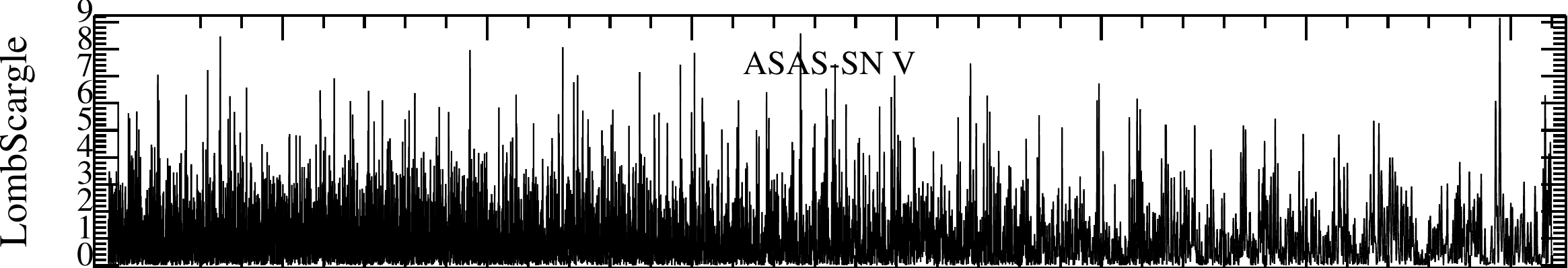}
\includegraphics[width=0.96\textwidth]{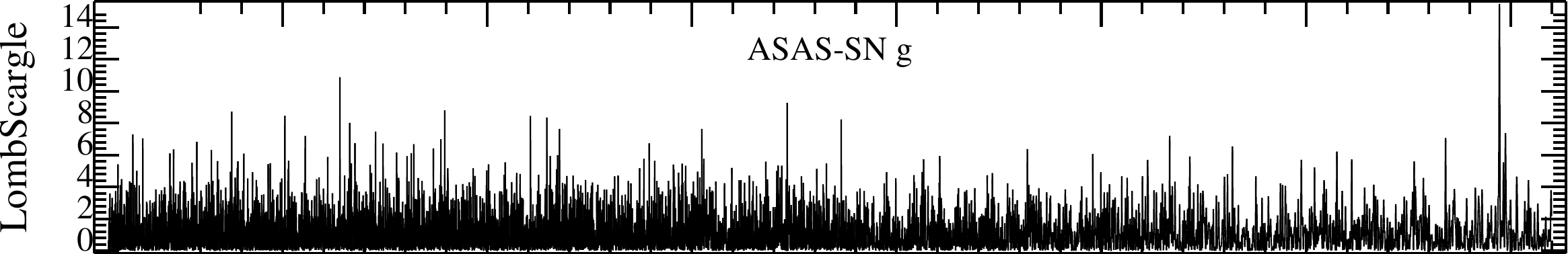}
\includegraphics[width=0.96\textwidth]{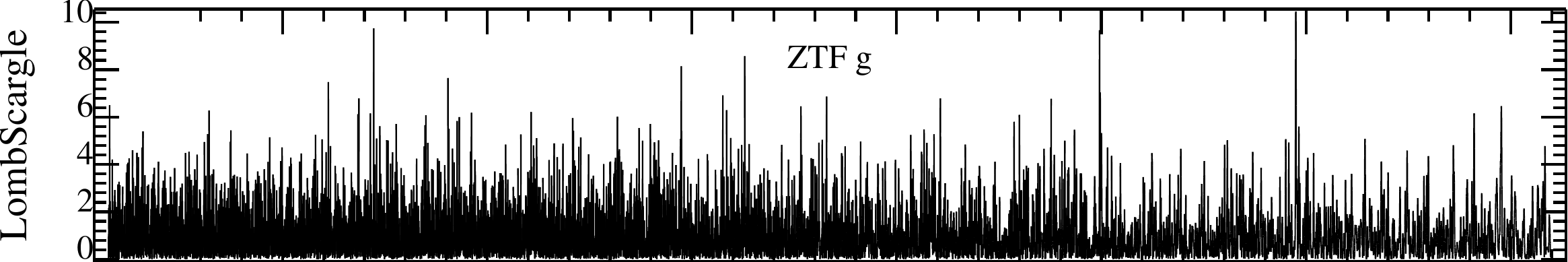}
\includegraphics[width=0.96\textwidth]{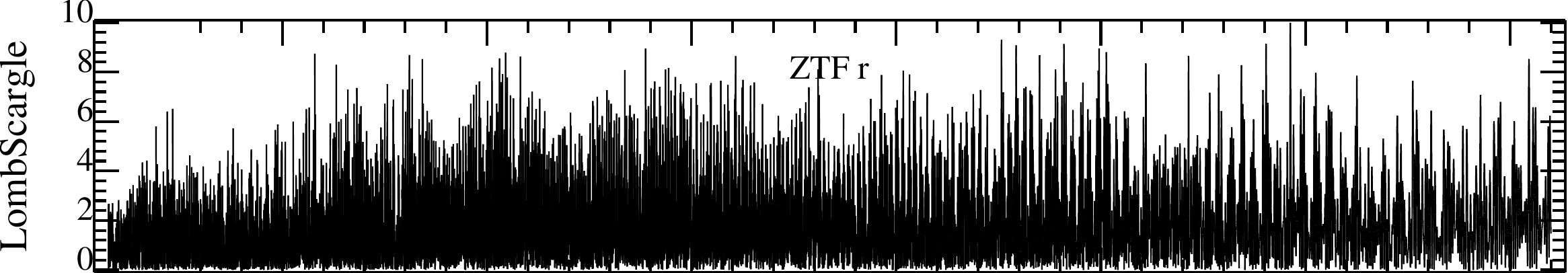}
\includegraphics[width=0.96\textwidth]{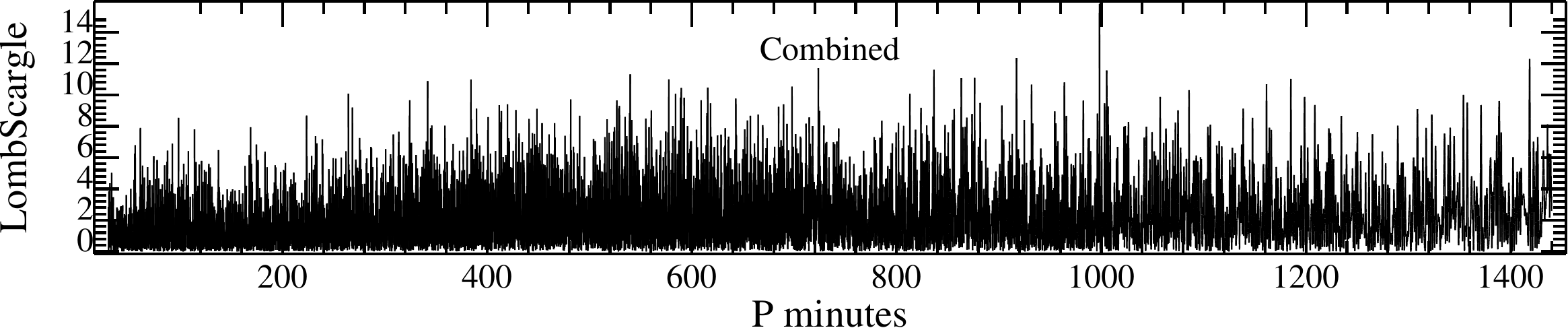}
\caption{
LombScargle results for ASAS-SN V, g band, ZTF g, r band light curves, 
and combined curves using \textbf{GATSPY},
from top to bottom, respectively.
}
\label{ztf_ls}
\end{figure*}

The Lomb-Scargle periodogram \citep{1976Ap&SS..39..447L, 1982ApJ...263..835S}
is used to find periods from ASAS-SN V, g band, ZTF g, r band light curves.
Also a combined search is processed with python toolkit \textbf{GATSPY} \citep[][]{VanderPlas2015Gatspy}.
The period ranges are from 30 minutes to 1 day, with a step of 6 seconds.
The results are shown in Figure \ref{ztf_ls}.
No clear period features are found.
We have also folded these light curves with the spectroscopic period but no orbital modulation can be identified.
The peaks around 1000, 1200 and 1400 minutes have been verified 
and all seem to be caused by sampling pattern.
Considering the light curves have $\sigma$ of 0.02mag, 
which are close to the instrument detection limits,
the actual orbital modulation amplitude should be less than 0.02 mag.

The Transiting Exoplanet Survey Satellite (TESS) \citep{tess} (Sector 18) have also covered the sky region of J0338, 
but unfortunately it is not within the Candidate Target List (CTL). 
Using \textbf{eleanor}: An Open-source Tool for Extracting Light Curves from the TESS Full-frame Image, a 30 minutes cadence light curve can be extracted.
The cadence is too long for the orbital period (about 1/6 phase)
and again no orbital modulation feature can be found.

In previous Sections, 
we have derived a radius of $\approx0.12$ $R_{\odot}$ and a filling factor of $\approx0.43$.
Using the Wilson-Devinney code \citep{Wilson1971WD},
we have simulated the light curves in V, g, r, i and TESS bands,
with inclination angles from $45^{\circ}$ to $90^{\circ}$.
The simulated light curves have peak to peak variation amplitudes in the range of $\sim0.8\%$ to $\sim1.6\%$.
Larger inclination angles correspond to larger amplitudes.
Such low amplitude variations can be verified 
by using larger telescope and more sensitive instruments.

%
%

If the orbital ephemeris is mandatorily chosen,
by fitting the folded light curves with cosine function, 
we derive semi-amplitudes of 0.55\%, 0.23\%, 0.82\%, 0.45\% and 0.24\%,
for ZTF g, r, ASAS-SN V, g and TESS bands, respectively.
These amplitudes suggest a moderate inclination angle (about 50$^\circ$).
However the uncertainties, using bootstrapping method, are on the order of few percents.

\section{Results and Discussion}
\label{sect:summary}

\begin{table*}
\caption{Summaries of parameters for J0338.
\label{tab.parameters}}
\setlength{\tabcolsep}{4.5pt}
\begin{center}
\begin{tabular}{cccc}
\hline\noalign{\smallskip}
Name & Description & Unit & Value \\
\hline\noalign{\smallskip}
RA & Right Ascension J2000 & [degrees] & 54.69611778 \\ 
DEC & Declination J2000 & [degrees]   & 41.57336115 \\ 
P & Orbital Period & [days] & 0.1253132(1) \\
T0 & Barycentric Julian Day & [days] & 2457034.9211(3) \\
V0 & System velocity & [km/s] & $-41\pm3$ \\
K1 & RV semi-amplitude & [km/s] & $289\pm4$ \\
\hline\noalign{\smallskip}
Plx & Parallax & [mas] &  $1.6775\pm0.0318$ \\
D & Distance & [pc] & $596\pm11$ \\
DM & Distance Modules & mag & $8.876\pm0.040$ \\
E(B-V) & \citet[][]{GreenDust2019} & [mag] & 0.16$\pm$ 0.02 \\
E(B-V) & \citet[][]{GreenDust2017} & [mag] & 0.12$\pm$ 0.02 \\
E(B-V) & \citet[][]{Schlegel1998} & [mag] & $\sim0.21$ \\
E(B-V) & \citet[][]{Planck2016Dustmap} & [mag] & $\sim0.24$ \\
\hline\noalign{\smallskip}
$T_{\rm eff}$ & Spectral temperature & [K] & 22575$\pm$494 \\
log $g$ & Spectral surface gravity & [dex] & 5.60$\pm$0.08 \\
log $[He/H]$ & Spectral He abundance & [-] & -3.87$\pm$0.57 \\
E(B-V) & SED fitted extinction & [mag] & $0.14\pm0.01$ \\
R1 & SED fitted radius & [$R_{\odot}$] & $0.121\pm0.003$ \\
$M1_{spec}$ & Star 1 spectral mass & [$M_{\odot}$] & $0.22\pm0.05$ \\
$M1_{model}$ & Star 1 model mass & [$M_{\odot}$] & $0.22\pm0.02$ \\
Age1 & Star 1 model age & [Myr] & $\sim350$ \\   
R1 & Star 1 model radius & [$R_{\odot}$] & $0.122\pm0.013$ \\
\hline\noalign{\smallskip}
$M2_{90}$ & Star 2 mass with inclination angle of 90$^{\circ}$  & [$M_{\odot}$] & $\approx0.60$ \\
$M2_{60.0}$ & Star 2 mass with inclination angle of 60.0$^{\circ}$ & [$M_{\odot}$] & $\approx0.79$ \\
$SMA$ & Orbital separation & [$R_{\odot}$] & $\sim1.1$ \\
$f1$ & Star 1 filling factor & [-] & $0.43\pm0.01$ \\ 
\hline\noalign{\smallskip}
$v\sin i$ & Projected rotation velocity & [km/s] & 44$\pm$12 \\
\hline\noalign{\smallskip}
\end{tabular}
\end{center}
\smallskip
\end{table*}

In this work, we have described the time domain spectroscopic observations and data analysis
for a binary system containing an ELM white dwarf and unseen white dwarf companion.
In all, 37 single exposure low resolution spectra are obtained from LAMOST and P200.
The atmosphere parameters are estimated to be $T_{\rm eff}\sim22500$ K and log $g\sim5.6$ dex.
SED fitting using parallax from Gaia eDR3, 
provide a distance of $\sim596pc$ and a radius of $0.121\pm0.003$ $R_{\odot}$.
The radius and spectral surface gravity corresponds to a mass of $0.22\pm0.05$ $M_{\odot}$,
which is consistent with the ELM evolutionary sequence estimation ($0.22\pm0.02$ $M_{\odot}$).
The RV measurements are used to derive the orbital parameters, 
with period $=0.1253132$ days,
semi-amplitude of star 1 K1$=289\pm4$ km/s,
system velocity $V_{sys} = -41\pm3$ km/s.
The ephemeris is also given with a zero point of $T0_{BJD}=2457034.9211$ days.

The orbital parameters correspond to binary mass function of 0.31 $M_{\odot}$.
Then the minimum mass for star 2 is 0.60 $M_{\odot}$.
If a median inclination of 60.0$^{\circ}$ is considered, the mass of the secondary could be $\approx0.79$ $M_{\odot}$.
No feature of the secondary component star is found in all the single exposure spectra.
A main sequence star with such mass would have visible flux contribution
in both the near infrared spectra and the SED,
so the companion is expected to be a white dwarf.
Assuming the temperature of the CO white dwarf is not higher than the visible component,
the flux contribution of the invisible component is less than 1\% of the visible one,
based on the radius ratio.

We have also checked the ROSAT \citep[][]{Boller2016ROSAT} and Swift \citep[][]{Evans2014Swift} data,
no X-ray signal is found for J0338.
With the HIgh-energy LIghtcurve GeneraTor \footnote{http://xmmuls.esac.esa.int/upperlimitserver/},
the upper limit at flux 0.2 to 2 keV range is $\sim3.6e-13$ and $\sim4.9e-13$ $erg/s/cm^2$,
according the ROSAT-SURVEY and XMM-NEWTON SLEW data.
The X-ray silence is consistent with the white dwarf nature.

\subsection{Is it a hot subdwarf?}


Hot subdwarfs and ELMs are close relatives in evolutionary process,
and their locations in the Hertzsprung–Russell diagram significantly overlap.
According to our stellar parameter estimation above (effective temperature and mass),
J0338 is more likely an ELM.
A hot subdwarf is generally hotter and more massive ($>0.3$ $M_{\odot}$) \citep[][]{Han2002subdwarf, Zhang2009, Heber2016subdwarf}.
We suspect that the over estimation on surface gravity by previous works
is caused by adopting a higher resolution (R$\sim1800$).
The resolutions in the blue band of single exposures are between 1200 and 1600.
The flux normalization of the continuous spectra also affects the result.
Follow up high S/N spectral observation may be helpful to better estimate the atmosphere parameters and hence the nature.

It is thought a hot subdwarf is burning helium in the core while an ELM is burning hydrogen in the envelope.
Currently only asteroseismology can probe the interior structure of a remote star.
\citet[][]{Bedding2011} discovered that 
pulsating red giants with hydrogen burning envelopes have gravity-mode period spacings of $\sim50s$,
while those with helium burning cores have longer period spacings of $\sim100$ to $300s$.
\citet[][]{Guo2018} suggested that
core helium burning and shell helium burning hot subdwarfs can be distinguished by
pulsation properties such as 
the rates of change of period of the p-mode pulsators, 
the numbers of mixed modes, 
and the period spacings of the g-mode pulsators. 
Thus pulsating might offer another way to distinguish hot subdwarfs and ELMs 
within the parameter overlapping region.
However, current light curves of J0338 show no pulsating features.
Neither the canonical ZZ Ceti instability strips \citep[][]{Gianninas2011, Gianninas2015, Corsico2016}
nor the hot subdwarf instability strip \citep[][]{Heber2016subdwarf}
covers the parameters of J0338 properly.
Currently only about twenty ELMs are known to be pulsating \citep[][]{Hermes2013, Gianninas2016, Zhang2016, Wang2020ELMPulsating}.
More photometric observations on larger ELM sample might be helpful to solve this issue more convincingly.

\subsection{Evolutionary status}

\begin{figure}
\center
\includegraphics[width=\columnwidth]{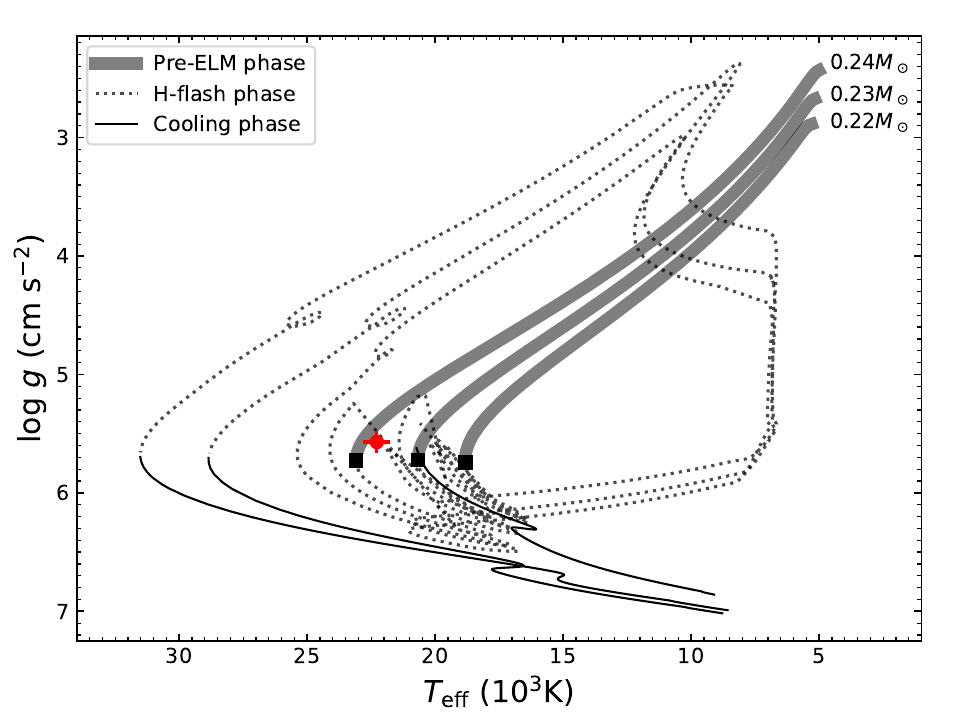}
\caption{
The observation parameters of J0338 in $\log T_{\rm eff}-\log g$ plane,
where the theoretical tracks are taken from \citet{Li2019ELM}. 
The pre-ELM phase is shown in thick lines.
The H-shell flash and cooling phases are shown in dotted lines and thin solid lines, respectively. 
The maximum temperature before H-shell flash is shown in black squares.}
\label{fig:tg}
\end{figure}

ELMs are believed to be formed within interaction binaries.
The hydrogen envelope of the progenitor is stripped by the companion
via either stable Roche lobe overflow (RL channel) or common envelope eject (CE channel) \citep[][]{Li2019ELM}. 
The 3 hours orbital period and $\sim0.22$ $M_{\odot}$ mass of J0338,
suggests that the system is probably formed via the CE channel,
according to the analysis of \citet[][]{Chen2017, Li2019ELM}.
ELMs with masses of $\sim0.22$ $M_{\odot}$ formed via RL channel are more likely to have orbital periods above one day.
In the CE channel,
the progenitor fills its Roche lobe during Hertzsprung gap (HG) or near the base of the red giant branch (RGB)
and the system enters CE stage due to the unstable mass transfer.
After the ejection of CE an low mass He-core ELM is formed with a lower mass limit of $\sim0.21$ $M_{\odot}$.

%

In Figure \ref{fig:tg}, we present the observation parameters of J0338 in $\log T_{\rm eff}-\log g$ plane, and the evolutionary tracks are taken from \citet{Li2019ELM}. After the termination of (unstable) mass transfer, the ELM WD enters into the contraction phase (also known as pre-ELM phase), where the luminosity is mainly supported by the hydrogen burning in the shell. The position of J0338 may suggest that the ELM WD is close to finishing the pre-ELM phase. The timescale of pre-ELM phase is strongly correlated with the ELM WD mass, and is about $1-5\times 10^8$ yr according to the results in \citet{Chen2017}. Since J0338 is supposed to be produced from common envelope ejection process and has well derived binary parameters, it can be used to constrain the common envelope ejection efficiency by reconstructing the evolutionary history \citep{zorotovic2010,scherbak2023}.

Due to the gravitational wave emission, 
the separation distance of the double white dwarf system will shrink and the two components will merge in the final end.
The time of merger of a double degenerate (DD) system can be estimated as follows \citep[][]{Chandra2021SDSS1337},
\begin{equation}
\tau_{GW}=10 \cdot \frac{(M1+M2)^{1/3}}{M1M2} P_{hours}^{8/3} Myr
\end{equation}.
With the orbital parameters of J0338, the merge time is $\sim1$ Gyr,
meaning the system will merge within a Hubble time.
As suggested by \citet[][]{Brown2016, Brown2020ELM},
J0338 will become a single massive white dwarf $\sim1$ $M_{\odot}$,
like most of the He+CO WD merging binaries.

Using the formula from \citet[][]{Kupfer2018},
J0338 has GW frequency of $\approx0.185$ mHz,
dimensionless GW amplitude of $\sim5e-23$,
and characteristic strain $h_c$ of about $\sim8e-21$,
assuming a 4 years observation time.
Unfortunately this is clearly below the LISA sensitivity curve
(see Figure 3 of \citet[][]{Kupfer2018}, and Figure 5 of \citet[][]{Chandra2021SDSS1337}). Nevertheless, the gravitational wave signal of J0338 will contribute to the foreground noise and exert an influence on the LISA sensitivity curve \citep[][]{lizw2020,Amaro2023LISA}.
Systems with 
shorter periods (from few 
minutes to about 1 hours) 
and smaller distance (few hundreds pc) are expected to
have strong S/N ratio \citep[][]{Brown2020ELM}.

\subsection{The projected rotation velocity $vsini$}
\label{sect:vsini}

Theoretically, due to the effect of tides on close binary, 
the rotation of each components will be synchronized with the orbit,
in a time depending on where the envelope is convective or radiative \citep[][]{Hurley2002Sync, Song2013Sync}.
Observational works also support synchronization, like \citet[][]{Wang2020ElCVn, Kim2021ElCVn}.
The synchronization time for J0338, assuming a convective layer, is less than $1$ Myr,
suggesting the ELM most probably in synchronization with orbit.
However, \citet[][]{Istrate2016ELMs} demonstrated that when the ELM contracts, 
the H envelope will rotate faster than orbit by up to 20 times.
This is explained by the angular momentum conservation of the H envelope,
and the speculation that He core of ELM isn't accelerated when H envelope contracts.

We have tried to measure the projected rotation velocity based on the low resolution spectra.
Due to the low resolution and low S/N, the uncertainties are naturally large.
Tests show that the spectral S/N of above 100 is required to obtain a good $vsini$ estimation.
Besides, due to integration effect caused by the relatively long exposure time, 
each observed single spectra is expected to be broadened by the RV variation, 
which is different at different orbital phase.
The resultant $vsini$ from single exposure fitting, using fixed atmosphere parameters, 
and subtracted by half the theoretic maximum RV variation so as to compensate the integration effect,
still varies from below 50km/s to above 100km/s.

To get higher S/N, the first 9 single exposures,
which have the same exposure length of 1500s, 
are corrected to the rest frame and merged into a single combined high S/N spectra.
Then the combined spectra is fitted under MCMC framework.
The resultant $vsini$ is $131\pm12$ km/s.
The square S/N weighted RV variations of each single exposure spectra,
calculated with the fitted sinusoidal RV curve from Section \ref{sect:orbit}, is $RVDiff\sim174$ km/s.
Thus the final $vsini$, after subtraction with half the RV variation ($RVDiff$), 
is $\sim44km$.
As the synchronized rotation velocity at $90^{\circ}$ is $51$ km/s,
based on the radius and orbital period of J0338 (Table \ref{tab.parameters}), 
the measured $vsini$ roughly suggests a synchronized rotation and an inclination of $\sim40^{\circ}$.
Nevertheless, the uncertainty of $RVDiff$, if estimated on the observation times of the 9 exposures,
is $\sim60$ km/s, indicating that uncertainty of $vsini$ will be comparable or even larger than the value of $vsini$.
While the combination process does increase the S/N ratio, 
the orbital integration effect is unfortunately enhanced at the same time,
due to phase differences between exposures.

\section*{acknowledgements}
Guoshoujing Telescope (the Large Sky Area Multi-Object Fiber Spectroscopic Telescope LAMOST) is a National Major Scientific Project built by the Chinese Academy of Sciences. Funding for the project has been provided by the National Development and Reform Commission. LAMOST is operated and managed by the National Astronomical Observatories, Chinese Academy of Sciences. 
This work presents results from the European Space Agency (ESA) space mission Gaia. Gaia data are being processed by the Gaia Data Processing and Analysis Consortium (DPAC). Funding for the DPAC is provided by national institutions, in particular the institutions participating in the Gaia MultiLateral Agreement (MLA). The Gaia mission website is https://www.cosmos.esa.int/gaia. The Gaia archive website is https://archives.esac.esa.int/gaia. We acknowledge use of the VizieR catalogue access tool, operated at CDS, Strasbourg, France, and of Astropy, a community-developed core Python package for Astronomy (Astropy Collaboration, 2013).
Z.H.T. would like to thank Fu Xiaoting, Zhang Xiaobin, Shao Yong, Liu Jifeng and Wangsong for the useful discussions and suggestions.
Z.H.T. thanks the support of the National Key R\&D Program of China (2019YFA0405000, 2022YFA1603002), NSFC 12090041, NSFC 11933004 and NSFC 12273056. 
Y.H.L. acknowledges support from the Youth Innovation Promotion Association of the CAS (Id. 2020060) and National Natural Science Foundation of China (Grant No. 11873066). 
ZWL and XFC thanks the support of National Key R\&D Program of China (Gant No. 2021YFA1600403), 
National Natural Science Foundation of China (Grant Nos. 11733008, 12103086, 12090040/12090043),
the National Science Fund for Distinguished Young Scholars (Grant No. 12125303)
and the Yunnan Fundamental Research Projects (No. 202101AU070276). 
We also acknowledge the science research grant from the China Manned Space Project with No.CMS-CSST-2021-A10.

\section*{Data Availability}
The data underlying this article will be shared on reasonable request to the corresponding author.



\bibliographystyle{mnras}
\bibliography{sample} 

\begin{thebibliography}{}
\makeatletter
\relax
\def\mn@urlcharsother{\let\do\@makeother \do\$\do\&\do\#\do\^\do\_\do\%\do\~}
\def\mn@doi{\begingroup\mn@urlcharsother \@ifnextchar [ {\mn@doi@}
  {\mn@doi@[]}}
\def\mn@doi@[#1]#2{\def\@tempa{#1}\ifx\@tempa\@empty \href
  {http://dx.doi.org/#2} {doi:#2}\else \href {http://dx.doi.org/#2} {#1}\fi
  \endgroup}
\def\mn@eprint#1#2{\mn@eprint@#1:#2::\@nil}
\def\mn@eprint@arXiv#1{\href {http://arxiv.org/abs/#1} {{\tt arXiv:#1}}}
\def\mn@eprint@dblp#1{\href {http://dblp.uni-trier.de/rec/bibtex/#1.xml}
  {dblp:#1}}
\def\mn@eprint@#1:#2:#3:#4\@nil{\def\@tempa {#1}\def\@tempb {#2}\def\@tempc
  {#3}\ifx \@tempc \@empty \let \@tempc \@tempb \let \@tempb \@tempa \fi \ifx
  \@tempb \@empty \def\@tempb {arXiv}\fi \@ifundefined
  {mn@eprint@\@tempb}{\@tempb:\@tempc}{\expandafter \expandafter \csname
  mn@eprint@\@tempb\endcsname \expandafter{\@tempc}}}

\bibitem[\protect\citeauthoryear{{Acernese} et~al.,}{{Acernese}
  et~al.}{2015}]{Acernese2015VIRGO}
{Acernese} F.,  et~al., 2015, \mn@doi [Classical and Quantum Gravity]
  {10.1088/0264-9381/32/2/024001}, \href
  {https://ui.adsabs.harvard.edu/abs/2015CQGra..32b4001A} {32, 024001}

\bibitem[\protect\citeauthoryear{{Ahumada} et~al.,}{{Ahumada}
  et~al.}{2020}]{2020ApJS..249....3A}
{Ahumada} R.,  et~al., 2020, \mn@doi [\apjs] {10.3847/1538-4365/ab929e}, \href
  {https://ui.adsabs.harvard.edu/abs/2020ApJS..249....3A} {249, 3}

\bibitem[\protect\citeauthoryear{{Althaus}, {Miller Bertolami}  \&
  {C{\'o}rsico}}{{Althaus} et~al.}{2013}]{Althaus2013ELM}
{Althaus} L.~G.,  {Miller Bertolami} M.~M.,   {C{\'o}rsico} A.~H.,  2013,
  \mn@doi [\aap] {10.1051/0004-6361/201321868}, \href
  {https://ui.adsabs.harvard.edu/abs/2013A&A...557A..19A} {557, A19}

\bibitem[\protect\citeauthoryear{{Amaro-Seoane} et~al.,}{{Amaro-Seoane}
  et~al.}{2023}]{Amaro2023LISA}
{Amaro-Seoane} P.,  et~al., 2023, \mn@doi [Living Reviews in Relativity]
  {10.1007/s41114-022-00041-y}, \href
  {https://ui.adsabs.harvard.edu/abs/2023LRR....26....2A} {26, 2}

\bibitem[\protect\citeauthoryear{{Badenes}, {Mullally}, {Thompson}  \&
  {Lupton}}{{Badenes} et~al.}{2009}]{Badenes2009SDSS1257}
{Badenes} C.,  {Mullally} F.,  {Thompson} S.~E.,   {Lupton} R.~H.,  2009,
  \mn@doi [\apj] {10.1088/0004-637X/707/2/971}, \href
  {https://ui.adsabs.harvard.edu/abs/2009ApJ...707..971B} {707, 971}

\bibitem[\protect\citeauthoryear{{Bai} et~al.,}{{Bai} et~al.}{2017}]{Bai2017}
{Bai} Z.,  et~al., 2017, \mn@doi [\pasp] {10.1088/1538-3873/129/972/024004},
  \href {https://ui.adsabs.harvard.edu/abs/2017PASP..129b4004B} {129, 024004}

\bibitem[\protect\citeauthoryear{{Bai} et~al.,}{{Bai} et~al.}{2021}]{Bai2021}
{Bai} Z.-R.,  et~al., 2021, \mn@doi [Research in Astronomy and Astrophysics]
  {10.1088/1674-4527/21/10/249}, \href
  {https://ui.adsabs.harvard.edu/abs/2021RAA....21..249B} {21, 249}

\bibitem[\protect\citeauthoryear{{Bedding} et~al.,}{{Bedding}
  et~al.}{2011}]{Bedding2011}
{Bedding} T.~R.,  et~al., 2011, \mn@doi [\nat] {10.1038/nature09935}, \href
  {https://ui.adsabs.harvard.edu/abs/2011Natur.471..608B} {471, 608}

\bibitem[\protect\citeauthoryear{{Bianchi}, {Shiao}  \& {Thilker}}{{Bianchi}
  et~al.}{2017}]{Bianchi2017galex}
{Bianchi} L.,  {Shiao} B.,   {Thilker} D.,  2017, \mn@doi [\apjs]
  {10.3847/1538-4365/aa7053}, \href
  {https://ui.adsabs.harvard.edu/abs/2017ApJS..230...24B} {230, 24}

\bibitem[\protect\citeauthoryear{{Boller}, {Freyberg}, {Tr{\"u}mper}, {Haberl},
  {Voges}  \& {Nandra}}{{Boller} et~al.}{2016}]{Boller2016ROSAT}
{Boller} T.,  {Freyberg} M.~J.,  {Tr{\"u}mper} J.,  {Haberl} F.,  {Voges} W.,
  {Nandra} K.,  2016, \mn@doi [\aap] {10.1051/0004-6361/201525648}, \href
  {https://ui.adsabs.harvard.edu/abs/2016A&A...588A.103B} {588, A103}

\bibitem[\protect\citeauthoryear{{Brown}, {Kilic}, {Brown}  \&
  {Kenyon}}{{Brown} et~al.}{2011}]{Brown2011}
{Brown} J.~M.,  {Kilic} M.,  {Brown} W.~R.,   {Kenyon} S.~J.,  2011, \mn@doi
  [\apj] {10.1088/0004-637X/730/2/67}, \href
  {https://ui.adsabs.harvard.edu/abs/2011ApJ...730...67B} {730, 67}

\bibitem[\protect\citeauthoryear{{Brown}, {Kilic}, {Kenyon}  \&
  {Gianninas}}{{Brown} et~al.}{2016}]{Brown2016}
{Brown} W.~R.,  {Kilic} M.,  {Kenyon} S.~J.,   {Gianninas} A.,  2016, \mn@doi
  [\apj] {10.3847/0004-637X/824/1/46}, \href
  {https://ui.adsabs.harvard.edu/abs/2016ApJ...824...46B} {824, 46}

\bibitem[\protect\citeauthoryear{{Brown} et~al.,}{{Brown}
  et~al.}{2020}]{Brown2020ELM}
{Brown} W.~R.,  et~al., 2020, \mn@doi [\apj] {10.3847/1538-4357/ab63cd}, \href
  {https://ui.adsabs.harvard.edu/abs/2020ApJ...889...49B} {889, 49}

\bibitem[\protect\citeauthoryear{{Brown}, {Kilic}, {Kosakowski}  \&
  {Gianninas}}{{Brown} et~al.}{2022}]{Brown2022ELM}
{Brown} W.~R.,  {Kilic} M.,  {Kosakowski} A.,   {Gianninas} A.,  2022, \mn@doi
  [\apj] {10.3847/1538-4357/ac72ac}, \href
  {https://ui.adsabs.harvard.edu/abs/2022ApJ...933...94B} {933, 94}

\bibitem[\protect\citeauthoryear{{Camarota} \& {Holberg}}{{Camarota} \&
  {Holberg}}{2014}]{Camarota2014Galex}
{Camarota} L.,  {Holberg} J.~B.,  2014, \mn@doi [\mnras]
  {10.1093/mnras/stt2422}, \href
  {https://ui.adsabs.harvard.edu/abs/2014MNRAS.438.3111C} {438, 3111}

\bibitem[\protect\citeauthoryear{{Chambers} et~al.,}{{Chambers}
  et~al.}{2016}]{2016arXiv161205560C}
{Chambers} K.~C.,  et~al., 2016, \mn@doi [arXiv e-prints]
  {10.48550/arXiv.1612.05560}, \href
  {https://ui.adsabs.harvard.edu/abs/2016arXiv161205560C} {p. arXiv:1612.05560}

\bibitem[\protect\citeauthoryear{{Chandra} et~al.,}{{Chandra}
  et~al.}{2021}]{Chandra2021SDSS1337}
{Chandra} V.,  et~al., 2021, \mn@doi [\apj] {10.3847/1538-4357/ac2145}, \href
  {https://ui.adsabs.harvard.edu/abs/2021ApJ...921..160C} {921, 160}

\bibitem[\protect\citeauthoryear{{Chen}, {Maxted}, {Li}  \& {Han}}{{Chen}
  et~al.}{2017}]{Chen2017}
{Chen} X.,  {Maxted} P.~F.~L.,  {Li} J.,   {Han} Z.,  2017, \mn@doi [\mnras]
  {10.1093/mnras/stx115}, \href
  {https://ui.adsabs.harvard.edu/abs/2017MNRAS.467.1874C} {467, 1874}

\bibitem[\protect\citeauthoryear{{C{\'o}rsico}, {Althaus}, {Serenelli},
  {Kepler}, {Jeffery}  \& {Corti}}{{C{\'o}rsico} et~al.}{2016}]{Corsico2016}
{C{\'o}rsico} A.~H.,  {Althaus} L.~G.,  {Serenelli} A.~M.,  {Kepler} S.~O.,
  {Jeffery} C.~S.,   {Corti} M.~A.,  2016, \mn@doi [\aap]
  {10.1051/0004-6361/201528032}, \href
  {https://ui.adsabs.harvard.edu/abs/2016A&A...588A..74C} {588, A74}

\bibitem[\protect\citeauthoryear{{Culpan}, {Geier}, {Reindl}, {Pelisoli},
  {Gentile Fusillo}  \& {Vorontseva}}{{Culpan}
  et~al.}{2022}]{Culpan2022subdwarf}
{Culpan} R.,  {Geier} S.,  {Reindl} N.,  {Pelisoli} I.,  {Gentile Fusillo} N.,
   {Vorontseva} A.,  2022, \mn@doi [\aap] {10.1051/0004-6361/202243337}, \href
  {https://ui.adsabs.harvard.edu/abs/2022A&A...662A..40C} {662, A40}

\bibitem[\protect\citeauthoryear{{Cutri} et~al.,}{{Cutri}
  et~al.}{2003}]{Cutri2003_2mass}
{Cutri} R.~M.,  et~al., 2003, VizieR Online Data Catalog, \href
  {https://ui.adsabs.harvard.edu/abs/2003yCat.2246....0C} {p. II/246}

\bibitem[\protect\citeauthoryear{{Cutri} et~al.,}{{Cutri}
  et~al.}{2021}]{Cutri2014_wise}
{Cutri} R.~M.,  et~al., 2021, VizieR Online Data Catalog, \href
  {https://ui.adsabs.harvard.edu/abs/2014yCat.2328....0C} {p. II/328}

\bibitem[\protect\citeauthoryear{{Eggleton}, {Fitchett}  \& {Tout}}{{Eggleton}
  et~al.}{1989}]{Eggleton1989}
{Eggleton} P.~P.,  {Fitchett} M.~J.,   {Tout} C.~A.,  1989, \mn@doi [\apj]
  {10.1086/168190}, \href
  {https://ui.adsabs.harvard.edu/abs/1989ApJ...347..998E} {347, 998}

\bibitem[\protect\citeauthoryear{{Evans} et~al.,}{{Evans}
  et~al.}{2014}]{Evans2014Swift}
{Evans} P.~A.,  et~al., 2014, \mn@doi [\apjs] {10.1088/0067-0049/210/1/8},
  \href {https://ui.adsabs.harvard.edu/abs/2014ApJS..210....8E} {210, 8}

\bibitem[\protect\citeauthoryear{{Gaia Collaboration} et~al.,}{{Gaia
  Collaboration} et~al.}{2018}]{2018gaia}
{Gaia Collaboration} et~al., 2018, \mn@doi [\aap]
  {10.1051/0004-6361/201833051}, \href
  {https://ui.adsabs.harvard.edu/abs/2018A&A...616A...1G} {616, A1}

\bibitem[\protect\citeauthoryear{{Gaia Collaboration} et~al.,}{{Gaia
  Collaboration} et~al.}{2021}]{2021Gaia}
{Gaia Collaboration} et~al., 2021, \mn@doi [\aap]
  {10.1051/0004-6361/202039657}, \href
  {https://ui.adsabs.harvard.edu/abs/2021A&A...649A...1G} {649, A1}

\bibitem[\protect\citeauthoryear{{Ge} et~al.,}{{Ge} et~al.}{2022}]{Ge2022CE}
{Ge} H.,  et~al., 2022, \mn@doi [\apj] {10.3847/1538-4357/ac75d3}, \href
  {https://ui.adsabs.harvard.edu/abs/2022ApJ...933..137G} {933, 137}

\bibitem[\protect\citeauthoryear{{Geier}, {Raddi}, {Gentile Fusillo}  \&
  {Marsh}}{{Geier} et~al.}{2019}]{Geier2019subdwarf}
{Geier} S.,  {Raddi} R.,  {Gentile Fusillo} N.~P.,   {Marsh} T.~R.,  2019,
  \mn@doi [\aap] {10.1051/0004-6361/201834236}, \href
  {https://ui.adsabs.harvard.edu/abs/2019A&A...621A..38G} {621, A38}

\bibitem[\protect\citeauthoryear{{Geier}, {Dorsch}, {Pelisoli}, {Reindl},
  {Heber}  \& {Irrgang}}{{Geier} et~al.}{2022}]{Geier2022}
{Geier} S.,  {Dorsch} M.,  {Pelisoli} I.,  {Reindl} N.,  {Heber} U.,
  {Irrgang} A.,  2022, \mn@doi [\aap] {10.1051/0004-6361/202143022}, \href
  {https://ui.adsabs.harvard.edu/abs/2022A&A...661A.113G} {661, A113}

\bibitem[\protect\citeauthoryear{{Gianninas}, {Bergeron}  \&
  {Ruiz}}{{Gianninas} et~al.}{2011}]{Gianninas2011}
{Gianninas} A.,  {Bergeron} P.,   {Ruiz} M.~T.,  2011, \mn@doi [\apj]
  {10.1088/0004-637X/743/2/138}, \href
  {https://ui.adsabs.harvard.edu/abs/2011ApJ...743..138G} {743, 138}

\bibitem[\protect\citeauthoryear{{Gianninas}, {Kilic}, {Brown}, {Canton}  \&
  {Kenyon}}{{Gianninas} et~al.}{2015}]{Gianninas2015}
{Gianninas} A.,  {Kilic} M.,  {Brown} W.~R.,  {Canton} P.,   {Kenyon} S.~J.,
  2015, \mn@doi [\apj] {10.1088/0004-637X/812/2/167}, \href
  {https://ui.adsabs.harvard.edu/abs/2015ApJ...812..167G} {812, 167}

\bibitem[\protect\citeauthoryear{{Gianninas}, {Curd}, {Fontaine}, {Brown}  \&
  {Kilic}}{{Gianninas} et~al.}{2016}]{Gianninas2016}
{Gianninas} A.,  {Curd} B.,  {Fontaine} G.,  {Brown} W.~R.,   {Kilic} M.,
  2016, \mn@doi [\apjl] {10.3847/2041-8205/822/2/L27}, \href
  {https://ui.adsabs.harvard.edu/abs/2016ApJ...822L..27G} {822, L27}

\bibitem[\protect\citeauthoryear{{Green} et~al.,}{{Green}
  et~al.}{2018}]{GreenDust2017}
{Green} G.~M.,  et~al., 2018, \mn@doi [\mnras] {10.1093/mnras/sty1008}, \href
  {https://ui.adsabs.harvard.edu/abs/2018MNRAS.478..651G} {478, 651}

\bibitem[\protect\citeauthoryear{{Green}, {Schlafly}, {Zucker}, {Speagle}  \&
  {Finkbeiner}}{{Green} et~al.}{2019}]{GreenDust2019}
{Green} G.~M.,  {Schlafly} E.,  {Zucker} C.,  {Speagle} J.~S.,   {Finkbeiner}
  D.,  2019, \mn@doi [\apj] {10.3847/1538-4357/ab5362}, \href
  {https://ui.adsabs.harvard.edu/abs/2019ApJ...887...93G} {887, 93}

\bibitem[\protect\citeauthoryear{{Grote} \& {LIGO Scientific
  Collaboration}}{{Grote} \& {LIGO Scientific Collaboration}}{2010}]{Grote2010}
{Grote} H.,  {LIGO Scientific Collaboration} 2010, \mn@doi [Classical and
  Quantum Gravity] {10.1088/0264-9381/27/8/084003}, \href
  {https://ui.adsabs.harvard.edu/abs/2010CQGra..27h4003G} {27, 084003}

\bibitem[\protect\citeauthoryear{{Guo}}{{Guo}}{2018}]{Guo2018}
{Guo} J.-J.,  2018, \mn@doi [\apj] {10.3847/1538-4357/aadd05}, \href
  {https://ui.adsabs.harvard.edu/abs/2018ApJ...866...58G} {866, 58}

\bibitem[\protect\citeauthoryear{{Han}, {Podsiadlowski}, {Maxted}, {Marsh}  \&
  {Ivanova}}{{Han} et~al.}{2002}]{Han2002subdwarf}
{Han} Z.,  {Podsiadlowski} P.,  {Maxted} P.~F.~L.,  {Marsh} T.~R.,   {Ivanova}
  N.,  2002, \mn@doi [\mnras] {10.1046/j.1365-8711.2002.05752.x}, \href
  {https://ui.adsabs.harvard.edu/abs/2002MNRAS.336..449H} {336, 449}

\bibitem[\protect\citeauthoryear{{Heber}}{{Heber}}{2016}]{Heber2016subdwarf}
{Heber} U.,  2016, \mn@doi [\pasp] {10.1088/1538-3873/128/966/082001}, \href
  {https://ui.adsabs.harvard.edu/abs/2016PASP..128h2001H} {128, 082001}

\bibitem[\protect\citeauthoryear{{Hermes} et~al.,}{{Hermes}
  et~al.}{2013}]{Hermes2013}
{Hermes} J.~J.,  et~al., 2013, \mn@doi [\mnras] {10.1093/mnras/stt1835}, \href
  {https://ui.adsabs.harvard.edu/abs/2013MNRAS.436.3573H} {436, 3573}

\bibitem[\protect\citeauthoryear{{Holberg}, {Oswalt}, {Sion}  \&
  {McCook}}{{Holberg} et~al.}{2016}]{Holberg2016}
{Holberg} J.~B.,  {Oswalt} T.~D.,  {Sion} E.~M.,   {McCook} G.~P.,  2016,
  \mn@doi [\mnras] {10.1093/mnras/stw1357}, \href
  {https://ui.adsabs.harvard.edu/abs/2016MNRAS.462.2295H} {462, 2295}

\bibitem[\protect\citeauthoryear{{Huang} et~al.,}{{Huang}
  et~al.}{2020}]{Huang2020TianQin}
{Huang} S.-J.,  et~al., 2020, \mn@doi [\prd] {10.1103/PhysRevD.102.063021},
  \href {https://ui.adsabs.harvard.edu/abs/2020PhRvD.102f3021H} {102, 063021}

\bibitem[\protect\citeauthoryear{{Hubeny}}{{Hubeny}}{1988}]{Hubeny1988TLUSTY}
{Hubeny} I.,  1988, \mn@doi [Computer Physics Communications]
  {10.1016/0010-4655(88)90177-4}, \href
  {https://ui.adsabs.harvard.edu/abs/1988CoPhC..52..103H} {52, 103}

\bibitem[\protect\citeauthoryear{{Hurley}, {Tout}  \& {Pols}}{{Hurley}
  et~al.}{2002}]{Hurley2002Sync}
{Hurley} J.~R.,  {Tout} C.~A.,   {Pols} O.~R.,  2002, \mn@doi [\mnras]
  {10.1046/j.1365-8711.2002.05038.x}, \href
  {https://ui.adsabs.harvard.edu/abs/2002MNRAS.329..897H} {329, 897}

\bibitem[\protect\citeauthoryear{{Istrate}, {Tauris}, {Langer}  \&
  {Antoniadis}}{{Istrate} et~al.}{2014}]{istrate2014}
{Istrate} A.~G.,  {Tauris} T.~M.,  {Langer} N.,   {Antoniadis} J.,  2014,
  \mn@doi [\aap] {10.1051/0004-6361/201424681}, \href
  {http://adsabs.harvard.edu/abs/2014A%26A...571L...3I} {571, L3}

\bibitem[\protect\citeauthoryear{{Istrate}, {Marchant}, {Tauris}, {Langer},
  {Stancliffe}  \& {Grassitelli}}{{Istrate} et~al.}{2016}]{Istrate2016ELMs}
{Istrate} A.~G.,  {Marchant} P.,  {Tauris} T.~M.,  {Langer} N.,  {Stancliffe}
  R.~J.,   {Grassitelli} L.,  2016, \mn@doi [\aap]
  {10.1051/0004-6361/201628874}, \href
  {https://ui.adsabs.harvard.edu/abs/2016A&A...595A..35I} {595, A35}

\bibitem[\protect\citeauthoryear{{Jim{\'e}nez-Esteban}, {Torres},
  {Rebassa-Mansergas}, {Cruz}, {Murillo-Ojeda}, {Solano}, {Rodrigo}  \&
  {Camisassa}}{{Jim{\'e}nez-Esteban} et~al.}{2023}]{Jim2023GaiaWD100pc3}
{Jim{\'e}nez-Esteban} F.~M.,  {Torres} S.,  {Rebassa-Mansergas} A.,  {Cruz} P.,
   {Murillo-Ojeda} R.,  {Solano} E.,  {Rodrigo} C.,   {Camisassa} M.~E.,  2023,
  \mn@doi [\mnras] {10.1093/mnras/stac3382}, \href
  {https://ui.adsabs.harvard.edu/abs/2023MNRAS.518.5106J} {518, 5106}

\bibitem[\protect\citeauthoryear{{Kagra Collaboration} et~al.,}{{Kagra
  Collaboration} et~al.}{2019}]{Kagra2019}
{Kagra Collaboration} et~al., 2019, \mn@doi [Nature Astronomy]
  {10.1038/s41550-018-0658-y}, \href
  {https://ui.adsabs.harvard.edu/abs/2019NatAs...3...35K} {3, 35}

\bibitem[\protect\citeauthoryear{{Kilic}, {Brown}, {Allende Prieto}, {Kenyon}
  \& {Panei}}{{Kilic} et~al.}{2010}]{Kilic2010}
{Kilic} M.,  {Brown} W.~R.,  {Allende Prieto} C.,  {Kenyon} S.~J.,   {Panei}
  J.~A.,  2010, \mn@doi [\apj] {10.1088/0004-637X/716/1/122}, \href
  {https://ui.adsabs.harvard.edu/abs/2010ApJ...716..122K} {716, 122}

\bibitem[\protect\citeauthoryear{{Kim} et~al.,}{{Kim}
  et~al.}{2021}]{Kim2021ElCVn}
{Kim} S.-L.,  et~al., 2021, \mn@doi [\aj] {10.3847/1538-3881/ac23de}, \href
  {https://ui.adsabs.harvard.edu/abs/2021AJ....162..212K} {162, 212}

\bibitem[\protect\citeauthoryear{{Kleinman} et~al.,}{{Kleinman}
  et~al.}{2013}]{Kleinman2013whitedwarf}
{Kleinman} S.~J.,  et~al., 2013, \mn@doi [\apjs] {10.1088/0067-0049/204/1/5},
  \href {https://ui.adsabs.harvard.edu/abs/2013ApJS..204....5K} {204, 5}

\bibitem[\protect\citeauthoryear{{Koester}, {Voss}, {Napiwotzki}, {Christlieb},
  {Homeier}, {Lisker}, {Reimers}  \& {Heber}}{{Koester}
  et~al.}{2009}]{Koester2009}
{Koester} D.,  {Voss} B.,  {Napiwotzki} R.,  {Christlieb} N.,  {Homeier} D.,
  {Lisker} T.,  {Reimers} D.,   {Heber} U.,  2009, \mn@doi [\aap]
  {10.1051/0004-6361/200912531}, \href
  {https://ui.adsabs.harvard.edu/abs/2009A&A...505..441K} {505, 441}

\bibitem[\protect\citeauthoryear{{Korol}, {Hallakoun}, {Toonen}  \&
  {Karnesis}}{{Korol} et~al.}{2022}]{Korol2022DD}
{Korol} V.,  {Hallakoun} N.,  {Toonen} S.,   {Karnesis} N.,  2022, \mn@doi
  [\mnras] {10.1093/mnras/stac415}, \href
  {https://ui.adsabs.harvard.edu/abs/2022MNRAS.511.5936K} {511, 5936}

\bibitem[\protect\citeauthoryear{{Kosakowski}, {Kilic}, {Brown}  \&
  {Gianninas}}{{Kosakowski} et~al.}{2020}]{Kosakowski2020ELM}
{Kosakowski} A.,  {Kilic} M.,  {Brown} W.~R.,   {Gianninas} A.,  2020, \mn@doi
  [\apj] {10.3847/1538-4357/ab8300}, \href
  {https://ui.adsabs.harvard.edu/abs/2020ApJ...894...53K} {894, 53}

\bibitem[\protect\citeauthoryear{{Kosakowski}, {Brown}, {Kilic}, {Kupfer},
  {B{\'e}dard}, {Gianninas}, {Ag{\"u}eros}  \& {Barrientos}}{{Kosakowski}
  et~al.}{2023}]{Kosakowski2023ELM}
{Kosakowski} A.,  {Brown} W.~R.,  {Kilic} M.,  {Kupfer} T.,  {B{\'e}dard} A.,
  {Gianninas} A.,  {Ag{\"u}eros} M.~A.,   {Barrientos} M.,  2023, \mn@doi
  [\apj] {10.3847/1538-4357/acd187}, \href
  {https://ui.adsabs.harvard.edu/abs/2023ApJ...950..141K} {950, 141}

\bibitem[\protect\citeauthoryear{{Kulkarni} \& {van Kerkwijk}}{{Kulkarni} \&
  {van Kerkwijk}}{2010}]{Kulkarni2010SDSS1257}
{Kulkarni} S.~R.,  {van Kerkwijk} M.~H.,  2010, \mn@doi [\apj]
  {10.1088/0004-637X/719/2/1123}, \href
  {https://ui.adsabs.harvard.edu/abs/2010ApJ...719.1123K} {719, 1123}

\bibitem[\protect\citeauthoryear{{Kupfer} et~al.,}{{Kupfer}
  et~al.}{2015}]{Kupfer2015sdB}
{Kupfer} T.,  et~al., 2015, \mn@doi [\aap] {10.1051/0004-6361/201425213}, \href
  {https://ui.adsabs.harvard.edu/abs/2015A&A...576A..44K} {576, A44}

\bibitem[\protect\citeauthoryear{{Kupfer} et~al.,}{{Kupfer}
  et~al.}{2018}]{Kupfer2018}
{Kupfer} T.,  et~al., 2018, \mn@doi [\mnras] {10.1093/mnras/sty1545}, \href
  {https://ui.adsabs.harvard.edu/abs/2018MNRAS.480..302K} {480, 302}

\bibitem[\protect\citeauthoryear{{Lei}, {Zhao}, {N{\'e}meth}  \& {Zhao}}{{Lei}
  et~al.}{2018}]{Lei2018sdB}
{Lei} Z.,  {Zhao} J.,  {N{\'e}meth} P.,   {Zhao} G.,  2018, \mn@doi [\apj]
  {10.3847/1538-4357/aae82b}, \href
  {https://ui.adsabs.harvard.edu/abs/2018ApJ...868...70L} {868, 70}

\bibitem[\protect\citeauthoryear{{Li}, {Chen}, {Chen}  \& {Han}}{{Li}
  et~al.}{2019}]{Li2019ELM}
{Li} Z.,  {Chen} X.,  {Chen} H.-L.,   {Han} Z.,  2019, \mn@doi [\apj]
  {10.3847/1538-4357/aaf9a1}, \href
  {https://ui.adsabs.harvard.edu/abs/2019ApJ...871..148L} {871, 148}

\bibitem[\protect\citeauthoryear{{Li}, {Chen}, {Chen}, {Li}, {Yu}  \&
  {Han}}{{Li} et~al.}{2020}]{lizw2020}
{Li} Z.,  {Chen} X.,  {Chen} H.-L.,  {Li} J.,  {Yu} S.,   {Han} Z.,  2020,
  \mn@doi [\apj] {10.3847/1538-4357/ab7dc2}, \href
  {https://ui.adsabs.harvard.edu/abs/2020ApJ...893....2L} {893, 2}

\bibitem[\protect\citeauthoryear{{Li}, {Chen}, {Ge}, {Chen}  \& {Han}}{{Li}
  et~al.}{2023}]{lizw2023}
{Li} Z.,  {Chen} X.,  {Ge} H.,  {Chen} H.-L.,   {Han} Z.,  2023, \mn@doi [\aap]
  {10.1051/0004-6361/202243893}, \href
  {https://ui.adsabs.harvard.edu/abs/2023A&A...669A..82L} {669, A82}

\bibitem[\protect\citeauthoryear{{Liebert}, {Bergeron}  \& {Holberg}}{{Liebert}
  et~al.}{2005}]{wdparameter}
{Liebert} J.,  {Bergeron} P.,   {Holberg} J.~B.,  2005, \mn@doi [\apjs]
  {10.1086/425738}, \href
  {https://ui.adsabs.harvard.edu/abs/2005ApJS..156...47L} {156, 47}

\bibitem[\protect\citeauthoryear{{Lindegren} et~al.,}{{Lindegren}
  et~al.}{2021}]{gaiaedr3plxzero}
{Lindegren} L.,  et~al., 2021, \mn@doi [\aap] {10.1051/0004-6361/202039653},
  \href {https://ui.adsabs.harvard.edu/abs/2021A&A...649A...4L} {649, A4}

\bibitem[\protect\citeauthoryear{{Liu}, {R{\"o}pke}  \& {Han}}{{Liu}
  et~al.}{2023}]{Liuz2023}
{Liu} Z.-W.,  {R{\"o}pke} F.~K.,   {Han} Z.,  2023, \mn@doi [Research in
  Astronomy and Astrophysics] {10.1088/1674-4527/acd89e}, \href
  {https://ui.adsabs.harvard.edu/abs/2023RAA....23h2001L} {23, 082001}

\bibitem[\protect\citeauthoryear{{Lomb}}{{Lomb}}{1976}]{1976Ap&SS..39..447L}
{Lomb} N.~R.,  1976, \mn@doi [\apss] {10.1007/BF00648343}, \href
  {https://ui.adsabs.harvard.edu/abs/1976Ap&SS..39..447L} {39, 447}

\bibitem[\protect\citeauthoryear{{Luo}, {N{\'e}meth}, {Liu}, {Deng}  \&
  {Han}}{{Luo} et~al.}{2016}]{Luo2016sdB}
{Luo} Y.-P.,  {N{\'e}meth} P.,  {Liu} C.,  {Deng} L.-C.,   {Han} Z.-W.,  2016,
  \mn@doi [\apj] {10.3847/0004-637X/818/2/202}, \href
  {https://ui.adsabs.harvard.edu/abs/2016ApJ...818..202L} {818, 202}

\bibitem[\protect\citeauthoryear{{Luo}, {N{\'e}meth}, {Wang}, {Wang}  \&
  {Han}}{{Luo} et~al.}{2021}]{Luo2021sdB}
{Luo} Y.,  {N{\'e}meth} P.,  {Wang} K.,  {Wang} X.,   {Han} Z.,  2021, \mn@doi
  [\apjs] {10.3847/1538-4365/ac11f6}, \href
  {https://ui.adsabs.harvard.edu/abs/2021ApJS..256...28L} {256, 28}

\bibitem[\protect\citeauthoryear{{McCleery} et~al.,}{{McCleery}
  et~al.}{2020}]{McCleery2020GaiaWD40pc2}
{McCleery} J.,  et~al., 2020, \mn@doi [\mnras] {10.1093/mnras/staa2030}, \href
  {https://ui.adsabs.harvard.edu/abs/2020MNRAS.499.1890M} {499, 1890}

\bibitem[\protect\citeauthoryear{{Monet} et~al.,}{{Monet}
  et~al.}{2003}]{usnob1.0}
{Monet} D.~G.,  et~al., 2003, \mn@doi [\aj] {10.1086/345888}, \href
  {https://ui.adsabs.harvard.edu/abs/2003AJ....125..984M} {125, 984}

\bibitem[\protect\citeauthoryear{{Morales-Rueda}, {Maxted}, {Marsh}, {North}
  \& {Heber}}{{Morales-Rueda} et~al.}{2003}]{Morales2003subdwarf}
{Morales-Rueda} L.,  {Maxted} P.~F.~L.,  {Marsh} T.~R.,  {North} R.~C.,
  {Heber} U.,  2003, \mn@doi [\mnras] {10.1046/j.1365-8711.2003.06088.x}, \href
  {https://ui.adsabs.harvard.edu/abs/2003MNRAS.338..752M} {338, 752}

\bibitem[\protect\citeauthoryear{{Morrissey} et~al.,}{{Morrissey}
  et~al.}{2007}]{Morrissey2007Galex}
{Morrissey} P.,  et~al., 2007, \mn@doi [\apjs] {10.1086/520512}, \href
  {https://ui.adsabs.harvard.edu/abs/2007ApJS..173..682M} {173, 682}

\bibitem[\protect\citeauthoryear{{Mullally}, {Badenes}, {Thompson}  \&
  {Lupton}}{{Mullally} et~al.}{2009}]{Mullally2009DD}
{Mullally} F.,  {Badenes} C.,  {Thompson} S.~E.,   {Lupton} R.,  2009, \mn@doi
  [\apjl] {10.1088/0004-637X/707/1/L51}, \href
  {https://ui.adsabs.harvard.edu/abs/2009ApJ...707L..51M} {707, L51}

\bibitem[\protect\citeauthoryear{{Napiwotzki}, {Karl}, {Lisker}, {Heber},
  {Christlieb}, {Reimers}, {Nelemans}  \& {Homeier}}{{Napiwotzki}
  et~al.}{2004}]{Napiwotzki2004}
{Napiwotzki} R.,  {Karl} C.~A.,  {Lisker} T.,  {Heber} U.,  {Christlieb} N.,
  {Reimers} D.,  {Nelemans} G.,   {Homeier} D.,  2004, \mn@doi [\apss]
  {10.1023/B:ASTR.0000044362.07416.6c}, \href
  {https://ui.adsabs.harvard.edu/abs/2004Ap&SS.291..321N} {291, 321}

\bibitem[\protect\citeauthoryear{{Napiwotzki} et~al.,}{{Napiwotzki}
  et~al.}{2020}]{Napiwotzki2020}
{Napiwotzki} R.,  et~al., 2020, \mn@doi [\aap] {10.1051/0004-6361/201629648},
  \href {https://ui.adsabs.harvard.edu/abs/2020A&A...638A.131N} {638, A131}

\bibitem[\protect\citeauthoryear{{Nelemans}}{{Nelemans}}{2010}]{Nelemans2010}
{Nelemans} G.,  2010, \mn@doi [\apss] {10.1007/s10509-010-0392-0}, \href
  {https://ui.adsabs.harvard.edu/abs/2010Ap&SS.329...25N} {329, 25}

\bibitem[\protect\citeauthoryear{{Nelemans} et~al.,}{{Nelemans}
  et~al.}{2005}]{Nelemans2005}
{Nelemans} G.,  et~al., 2005, \mn@doi [\aap] {10.1051/0004-6361:20053174},
  \href {https://ui.adsabs.harvard.edu/abs/2005A&A...440.1087N} {440, 1087}

\bibitem[\protect\citeauthoryear{{Nemeth}, {{\"O}stensen}, {Tremblay}  \&
  {Hubeny}}{{Nemeth} et~al.}{2014}]{2014ASPC..481...95N}
{Nemeth} P.,  {{\"O}stensen} R.,  {Tremblay} P.,   {Hubeny} I.,  2014, in {van
  Grootel} V.,  {Green} E.,  {Fontaine} G.,   {Charpinet} S.,  eds,
  Astronomical Society of the Pacific Conference Series Vol. 481, 6th Meeting
  on Hot Subdwarf Stars and Related Objects. p.~95 (\mn@eprint {arXiv}
  {1308.0252})

\bibitem[\protect\citeauthoryear{{O'Brien} et~al.,}{{O'Brien}
  et~al.}{2023}]{OBrien2023GaiaWD40pc3}
{O'Brien} M.~W.,  et~al., 2023, \mn@doi [\mnras] {10.1093/mnras/stac3303},
  \href {https://ui.adsabs.harvard.edu/abs/2023MNRAS.518.3055O} {518, 3055}

\bibitem[\protect\citeauthoryear{{{\O}stensen} \& {van Winckel}}{{{\O}stensen}
  \& {van Winckel}}{2011}]{Ostensen2011}
{{\O}stensen} R.~H.,  {van Winckel} H.,  2011, in {Schmidtobreick} L.,
  {Schreiber} M.~R.,   {Tappert} C.,  eds,  Astronomical Society of the Pacific
  Conference Series Vol. 447, Evolution of Compact Binaries. p.~171

\bibitem[\protect\citeauthoryear{{Pelisoli} \& {Vos}}{{Pelisoli} \&
  {Vos}}{2019}]{Pelisoli2019GaiaELM}
{Pelisoli} I.,  {Vos} J.,  2019, \mn@doi [\mnras] {10.1093/mnras/stz1876},
  \href {https://ui.adsabs.harvard.edu/abs/2019MNRAS.488.2892P} {488, 2892}

\bibitem[\protect\citeauthoryear{{Planck Collaboration} et~al.,}{{Planck
  Collaboration} et~al.}{2016}]{Planck2016Dustmap}
{Planck Collaboration} et~al., 2016, \mn@doi [\aap]
  {10.1051/0004-6361/201629022}, \href
  {https://ui.adsabs.harvard.edu/abs/2016A&A...596A.109P} {596, A109}

\bibitem[\protect\citeauthoryear{{Price-Whelan}, {Hogg}, {Foreman-Mackey}  \&
  {Rix}}{{Price-Whelan} et~al.}{2017}]{2017ApJ...837...20P}
{Price-Whelan} A.~M.,  {Hogg} D.~W.,  {Foreman-Mackey} D.,   {Rix} H.-W.,
  2017, \mn@doi [\apj] {10.3847/1538-4357/aa5e50}, \href
  {https://ui.adsabs.harvard.edu/abs/2017ApJ...837...20P} {837, 20}

\bibitem[\protect\citeauthoryear{{Ram{\'\i}rez}, {Allende Prieto}  \&
  {Lambert}}{{Ram{\'\i}rez} et~al.}{2013}]{2013ApJ...764...78R}
{Ram{\'\i}rez} I.,  {Allende Prieto} C.,   {Lambert} D.~L.,  2013, \mn@doi
  [\apj] {10.1088/0004-637X/764/1/78}, \href
  {https://ui.adsabs.harvard.edu/abs/2013ApJ...764...78R} {764, 78}

\bibitem[\protect\citeauthoryear{{Ricker} et~al.,}{{Ricker}
  et~al.}{2015}]{tess}
{Ricker} G.~R.,  et~al., 2015, \mn@doi [Journal of Astronomical Telescopes,
  Instruments, and Systems] {10.1117/1.JATIS.1.1.014003}, \href
  {https://ui.adsabs.harvard.edu/abs/2015JATIS...1a4003R} {1, 014003}

\bibitem[\protect\citeauthoryear{{Saffer}, {Livio}  \& {Yungelson}}{{Saffer}
  et~al.}{1998}]{Saffer1998}
{Saffer} R.~A.,  {Livio} M.,   {Yungelson} L.~R.,  1998, \mn@doi [\apj]
  {10.1086/305907}, \href
  {https://ui.adsabs.harvard.edu/abs/1998ApJ...502..394S} {502, 394}

\bibitem[\protect\citeauthoryear{{Scargle}}{{Scargle}}{1982}]{1982ApJ...263..835S}
{Scargle} J.~D.,  1982, \mn@doi [\apj] {10.1086/160554}, \href
  {https://ui.adsabs.harvard.edu/abs/1982ApJ...263..835S} {263, 835}

\bibitem[\protect\citeauthoryear{{Scherbak} \& {Fuller}}{{Scherbak} \&
  {Fuller}}{2023}]{scherbak2023}
{Scherbak} P.,  {Fuller} J.,  2023, \mn@doi [\mnras] {10.1093/mnras/stac3313},
  \href {https://ui.adsabs.harvard.edu/abs/2023MNRAS.518.3966S} {518, 3966}

\bibitem[\protect\citeauthoryear{{Schlegel}, {Finkbeiner}  \&
  {Davis}}{{Schlegel} et~al.}{1998}]{Schlegel1998}
{Schlegel} D.~J.,  {Finkbeiner} D.~P.,   {Davis} M.,  1998, \mn@doi [\apj]
  {10.1086/305772}, \href
  {https://ui.adsabs.harvard.edu/abs/1998ApJ...500..525S} {500, 525}

\bibitem[\protect\citeauthoryear{{Solheim}}{{Solheim}}{2010}]{Solheim2010AmCVn}
{Solheim} J.~E.,  2010, \mn@doi [\pasp] {10.1086/656680}, \href
  {https://ui.adsabs.harvard.edu/abs/2010PASP..122.1133S} {122, 1133}

\bibitem[\protect\citeauthoryear{{Song}, {Maeder}, {Meynet}, {Huang},
  {Ekstr{\"o}m}  \& {Granada}}{{Song} et~al.}{2013}]{Song2013Sync}
{Song} H.~F.,  {Maeder} A.,  {Meynet} G.,  {Huang} R.~Q.,  {Ekstr{\"o}m} S.,
  {Granada} A.,  2013, \mn@doi [\aap] {10.1051/0004-6361/201321870}, \href
  {https://ui.adsabs.harvard.edu/abs/2013A&A...556A.100S} {556, A100}

\bibitem[\protect\citeauthoryear{{Su} \& {Cui}}{{Su} \&
  {Cui}}{2004}]{2004ChJAA...4....1S}
{Su} D.-Q.,  {Cui} X.-Q.,  2004, \mn@doi [\cjaa] {10.1088/1009-9271/4/1/1},
  \href {https://ui.adsabs.harvard.edu/abs/2004ChJAA...4....1S} {4, 1}

\bibitem[\protect\citeauthoryear{{Torres}, {Cantero}, {Rebassa-Mansergas},
  {Skorobogatov}, {Jim{\'e}nez-Esteban}  \& {Solano}}{{Torres}
  et~al.}{2019}]{Torres2019GaiaWD100pc1}
{Torres} S.,  {Cantero} C.,  {Rebassa-Mansergas} A.,  {Skorobogatov} G.,
  {Jim{\'e}nez-Esteban} F.~M.,   {Solano} E.,  2019, \mn@doi [\mnras]
  {10.1093/mnras/stz814}, \href
  {https://ui.adsabs.harvard.edu/abs/2019MNRAS.485.5573T} {485, 5573}

\bibitem[\protect\citeauthoryear{{Torres}, {Canals}, {Jim{\'e}nez-Esteban},
  {Rebassa-Mansergas}  \& {Solano}}{{Torres}
  et~al.}{2022}]{Torres2022GaiaWD100pc2}
{Torres} S.,  {Canals} P.,  {Jim{\'e}nez-Esteban} F.~M.,  {Rebassa-Mansergas}
  A.,   {Solano} E.,  2022, \mn@doi [\mnras] {10.1093/mnras/stac374}, \href
  {https://ui.adsabs.harvard.edu/abs/2022MNRAS.511.5462T} {511, 5462}

\bibitem[\protect\citeauthoryear{{Tremblay} et~al.,}{{Tremblay}
  et~al.}{2020}]{Tremblay2020GaiaWD40pc1}
{Tremblay} P.~E.,  et~al., 2020, \mn@doi [\mnras] {10.1093/mnras/staa1892},
  \href {https://ui.adsabs.harvard.edu/abs/2020MNRAS.497..130T} {497, 130}

\bibitem[\protect\citeauthoryear{{VanderPlas} \& {Ivezi{\'c}}}{{VanderPlas} \&
  {Ivezi{\'c}}}{2015}]{VanderPlas2015Gatspy}
{VanderPlas} J.~T.,  {Ivezi{\'c}} {\v{Z}}.,  2015, \mn@doi [\apj]
  {10.1088/0004-637X/812/1/18}, \href
  {https://ui.adsabs.harvard.edu/abs/2015ApJ...812...18V} {812, 18}

\bibitem[\protect\citeauthoryear{{Vos}, {Vu{\v{c}}kovi{\'c}}, {Chen}, {Han},
  {Boudreaux}, {Barlow}, {{\O}stensen}  \& {N{\'e}meth}}{{Vos}
  et~al.}{2019}]{Vos2019RL}
{Vos} J.,  {Vu{\v{c}}kovi{\'c}} M.,  {Chen} X.,  {Han} Z.,  {Boudreaux} T.,
  {Barlow} B.~N.,  {{\O}stensen} R.,   {N{\'e}meth} P.,  2019, Contributions of
  the Astronomical Observatory Skalnate Pleso, \href
  {https://ui.adsabs.harvard.edu/abs/2019CoSka..49..264V} {49, 264}

\bibitem[\protect\citeauthoryear{{Wall}, {Kilic}, {Bergeron}, {Rolland},
  {Genest-Beaulieu}  \& {Gianninas}}{{Wall} et~al.}{2019}]{Wall2019Galex}
{Wall} R.~E.,  {Kilic} M.,  {Bergeron} P.,  {Rolland} B.,  {Genest-Beaulieu}
  C.,   {Gianninas} A.,  2019, \mn@doi [\mnras] {10.1093/mnras/stz2506}, \href
  {https://ui.adsabs.harvard.edu/abs/2019MNRAS.489.5046W} {489, 5046}

\bibitem[\protect\citeauthoryear{{Wang} \& {Han}}{{Wang} \&
  {Han}}{2012}]{Wang2012IaSN}
{Wang} B.,  {Han} Z.,  2012, \mn@doi [\nar] {10.1016/j.newar.2012.04.001},
  \href {https://ui.adsabs.harvard.edu/abs/2012NewAR..56..122W} {56, 122}

\bibitem[\protect\citeauthoryear{{Wang}, {Su}, {Chu}, {Cui}  \& {Wang}}{{Wang}
  et~al.}{1996}]{1996ApOpt..35.5155W}
{Wang} S.-G.,  {Su} D.-Q.,  {Chu} Y.-Q.,  {Cui} X.,   {Wang} Y.-N.,  1996,
  \mn@doi [\ao] {10.1364/AO.35.005155}, \href
  {https://ui.adsabs.harvard.edu/abs/1996ApOpt..35.5155W} {35, 5155}

\bibitem[\protect\citeauthoryear{{Wang} et~al.,}{{Wang}
  et~al.}{2020a}]{Wang2020ElCVn}
{Wang} L.,  et~al., 2020a, \mn@doi [\aj] {10.3847/1538-3881/ab52fa}, \href
  {https://ui.adsabs.harvard.edu/abs/2020AJ....159....4W} {159, 4}

\bibitem[\protect\citeauthoryear{{Wang}, {Zhang}  \& {Dai}}{{Wang}
  et~al.}{2020b}]{Wang2020ELMPulsating}
{Wang} K.,  {Zhang} X.,   {Dai} M.,  2020b, \mn@doi [\apj]
  {10.3847/1538-4357/ab584c}, \href
  {https://ui.adsabs.harvard.edu/abs/2020ApJ...888...49W} {888, 49}

\bibitem[\protect\citeauthoryear{{Wang}, {N{\'e}meth}, {Luo}, {Chen}, {Jiang}
  \& {Cao}}{{Wang} et~al.}{2022}]{Wang2022ELM}
{Wang} K.,  {N{\'e}meth} P.,  {Luo} Y.,  {Chen} X.,  {Jiang} Q.,   {Cao} X.,
  2022, \mn@doi [\apj] {10.3847/1538-4357/ac847c}, \href
  {https://ui.adsabs.harvard.edu/abs/2022ApJ...936....5W} {936, 5}

\bibitem[\protect\citeauthoryear{{Werner}, {Dreizler}  \& {Rauch}}{{Werner}
  et~al.}{2012}]{TMAP}
{Werner} K.,  {Dreizler} S.,   {Rauch} T.,  2012, {TMAP: T{\"u}bingen NLTE
  Model-Atmosphere Package}, Astrophysics Source Code Library, record
  ascl:1212.015 (\mn@eprint {ascl} {1212.015})

\bibitem[\protect\citeauthoryear{{Wilson} \& {Devinney}}{{Wilson} \&
  {Devinney}}{1971}]{Wilson1971WD}
{Wilson} R.~E.,  {Devinney} E.~J.,  1971, \mn@doi [\apj] {10.1086/150986},
  \href {https://ui.adsabs.harvard.edu/abs/1971ApJ...166..605W} {166, 605}

\bibitem[\protect\citeauthoryear{{Yuan} et~al.,}{{Yuan}
  et~al.}{2023}]{Yuan2023}
{Yuan} H.,  et~al., 2023, \mn@doi [\aj] {10.3847/1538-3881/acaf07}, \href
  {https://ui.adsabs.harvard.edu/abs/2023AJ....165..119Y} {165, 119}

\bibitem[\protect\citeauthoryear{{Zacharias}, {Finch}, {Girard}, {Henden},
  {Bartlett}, {Monet}  \& {Zacharias}}{{Zacharias} et~al.}{2013}]{ucac4}
{Zacharias} N.,  {Finch} C.~T.,  {Girard} T.~M.,  {Henden} A.,  {Bartlett}
  J.~L.,  {Monet} D.~G.,   {Zacharias} M.~I.,  2013, \mn@doi [\aj]
  {10.1088/0004-6256/145/2/44}, \href
  {https://ui.adsabs.harvard.edu/abs/2013AJ....145...44Z} {145, 44}

\bibitem[\protect\citeauthoryear{{Zhang}, {Chen}  \& {Han}}{{Zhang}
  et~al.}{2009}]{Zhang2009}
{Zhang} X.,  {Chen} X.,   {Han} Z.,  2009, \mn@doi [\aap]
  {10.1051/0004-6361/200912336}, \href
  {https://ui.adsabs.harvard.edu/abs/2009A&A...504L..13Z} {504, L13}

\bibitem[\protect\citeauthoryear{{Zhang}, {Fu}, {Li}, {Ren}  \& {Luo}}{{Zhang}
  et~al.}{2016}]{Zhang2016}
{Zhang} X.~B.,  {Fu} J.~N.,  {Li} Y.,  {Ren} A.~B.,   {Luo} C.~Q.,  2016,
  \mn@doi [\apjl] {10.3847/2041-8205/821/2/L32}, \href
  {https://ui.adsabs.harvard.edu/abs/2016ApJ...821L..32Z} {821, L32}

\bibitem[\protect\citeauthoryear{{Zhong}, {Han}, {Luo}  \& {Wu}}{{Zhong}
  et~al.}{2023}]{Zhong2023Taiji}
{Zhong} X.,  {Han} W.-B.,  {Luo} Z.,   {Wu} Y.,  2023, \mn@doi [Science China
  Physics, Mechanics, and Astronomy] {10.1007/s11433-022-2028-7}, \href
  {https://ui.adsabs.harvard.edu/abs/2023SCPMA..6630411Z} {66, 230411}

\bibitem[\protect\citeauthoryear{{Zorotovic}, {Schreiber}, {G{\"a}nsicke}  \&
  {Nebot G{\'o}mez-Mor{\'a}n}}{{Zorotovic} et~al.}{2010}]{zorotovic2010}
{Zorotovic} M.,  {Schreiber} M.~R.,  {G{\"a}nsicke} B.~T.,   {Nebot
  G{\'o}mez-Mor{\'a}n} A.,  2010, \mn@doi [\aap] {10.1051/0004-6361/200913658},
  \href {https://ui.adsabs.harvard.edu/abs/2010A&A...520A..86Z} {520, A86}

\makeatother
\end{thebibliography}



\bsp	
\label{lastpage}
\end{document}